 \documentclass{emulateapj}

\usepackage{graphicx}
\usepackage{epsfig}
\usepackage{times}
\usepackage{natbib}
\usepackage{url}
\usepackage{color}
\usepackage{amssymb,amsmath}
\usepackage{hhline}

\newcommand\mb{\mathbf}

\shorttitle{Magnetic Reconnection} \shortauthors{Leake et al.}

\begin{document}

\title{The Onset of 3D Magnetic Reconnection and Heating in the Solar Corona}

\author{
J.~E. Leake$^{1}$,
L.~K.~S. Daldorff$^{2}$,
and J.~A. Klimchuk$^{1}$} 

\affil{$1$ NASA Goddard Space Flight Center, Greenbelt MD, USA}
\affil{$2$ Catholic University of America, Washington DC, USA}

\begin{abstract}

Magnetic reconnection, a fundamentally important process in many aspects of heliophysics and astrophysics, is believed to be initiated by the tearing instability of an \textit{electric current sheet}, a region where magnetic field abruptly changes direction and electric currents build up. Recent studies have suggested that the amount of magnetic shear in these structures is a critical parameter for the \textit{switch-on} nature of magnetic reconnection in the solar atmosphere, at fluid spatial scales much larger than kinetic scales. We present results of visco-resistive magnetohydrodynamic (MHD) simulations of magnetic reconnection in 3D current sheets with conditions appropriate to the solar corona. Using high-fidelity simulations, we follow the evolution of the linear and non-linear 3D tearing instability, leading to magnetic reconnection, with analysis of the evolution of parallel and oblique modes of this instability. We find that, depending on the parameter space, magnetic shear can play a vital role in the onset of significant energy release and heating via non-linear tearing. 
Two regimes in our study exist, dependent on whether the current sheet is longer or shorter than the wavelength of the fastest growing parallel mode (in the corresponding infinite system), thus determining whether sub-harmonics are present in the actual system. In one parameter regime, where the fastest growing parallel mode has sub-harmonics, the non-linear interaction of these sub-harmonics and the subsequent coalescence of 3D plasmoids dominates the non-linear evolution, with magnetic shear playing only a weak role in the amount of energy released. In the second parameter regime, where the fastest growing parallel mode has no-sub-harmonics, then only strongly sheared current sheets, where oblique mode are strong enough to compete with the dominant parallel mode, show any significant energy release. We expect both regimes to exist on the Sun, and so our results have important consequences for the the question of reconnection onset in different solar physics applications.

\end{abstract}


\section{Introduction}
\label{sec:intro}
Magnetic reconnection, the process which converts energy stored in magnetic fields into kinetic and thermal energy of the plasma by breaking magnetic connectivity, is observed in a range of laboratory and astrophysical plasmas. In solar physics, magnetic reconnection is believed to be the cause of transient phenomena such as solar flares, X-ray jets, and plays a significant role in coronal mass ejections (CMEs), as well as playing a crucial role in the heating of the atmosphere itself \citep[e.g.][]{Karpen_2012, Wyper_2017, Klimchuk_2015}.

In both the triggering of dynamic events and the heating of the solar atmosphere, the \textit{switch-on} nature of the process is key. The energy is built up over long timescales, due to a combination of the emergence of complex magnetic fields from the interior and multi-scale solar surface motions injecting further complexity into these magnetic fields, and released on short timescales, indicating some kind of threshold. At a certain, as yet, unspecified limit, complex magnetic fields will violently reconnect to reduce this complexity, and in the process expel mass, magnetic flux, and energetic particles. In the context of coronal heating, it is believed that the heating is temporally sporadic, with the timescale of energy release due to magnetic reconnection  being shorter than the cooling time. If magnetic reconnection were ubiquitously efficient, there would not be enough time to build up the magnetic energy required to power CMEs and flares, and coronal heating would be too weak to explain the observed temperatures \citep{Klimchuk_2015}. Furthermore, the properties of observed coronal emission, that of systematic delays between emissions at different temperatures, and the presence of very high ($>$ 5 MK) temperatures, suggest that steady heating is not a viable situation in the corona \citep{Klimchuk_2019}. Fortuitously, magnetic reconnection in the nearly perfectly conducting solar corona is typically very weak (see below). Thus, an important question to answer is: how does magnetic reconnection rapidly switch on in the solar corona?

The original work of Sweet and Parker \citep{Sweet_1958,Parker_1957} formulated steady-state magnetic reconnection as a local fluid process occurring in a current layer of width $\delta$ much smaller than its length $L$, and plasma with non-zero resistivity ($\eta$) due to electron-ion collisions, and Lundquist number $S=\mu_0LV_{a}/\eta$ with $V_{a}$ being the local Alfv\'{e}n speed. The reconnection rate and aspect ratio ($\delta/L$) both scaled with $S^{-1/2}$. In the corona, $S \sim 10^{11}$ which suggests that the Sweet-Parker width is  comparable to kinetic scales, and, furthermore, that the reconnection rate is too slow to explain the observed solar phenomena mentioned above.

A complicating issue related to magnetic reconnection in the corona is that the dynamic length scales which dominate the magnetic energy build up process are in the fluid regime.
However, the breaking of magnetic fieldlines occurs on kinetic scales. This has led to a practical barrier for advancement. When it comes to modeling dynamic events in the solar corona, the scales of energy release are too small to resolve, and detailed studies of collisionless reconnection cannot cover the length-scales to model the interaction of kinetic reconnection and coronal scales. 

The  work of \citet{Loureiro_2007,Huang_2013}  and others showed that the steady-state Sweet-Parker reconnecting current sheet is unstable to the tearing (or plasmoid) instability, leading to fast, non-laminar reconnection on the macroscopic scales.
More recently, it has been demonstrated that a thinning current sheet will exhibit fast tearing at scales much larger than the Sweet-Parker width \citep{Pucci_2014,Del_Zanna_2016,Pucci_2017,Tenerani_2016}.
This suggests current sheet thickness is a key parameter in the desired \textit{switch-on} nature of magnetic reconnection in the corona.

 \citet{Dahlburg_1992} showed that the flux tubes formed by the tearing instability are susceptible to kinking, which leads to a more virulent release of energy. This was termed secondary instability. \cite{Dahlburg_2005,Dahlburg_2009,Dahlburg_2006} subsequently found that the kinking is stabilized by the presence of a guide field. They identified a critical shear angle (the angle through which the field rotates across the current sheet), which must be exceeded to get a large release of energy.


It appears that the fundamental magnetohydrodynamic (MHD) parameters of current-sheet thickness and magnetic shear  could both play critical roles in the onset of magnetic reconnection in the solar
corona. To date, a MHD 3D study of the tearing  instability in the corona, covering a self-consistent evolution of both parallel and oblique modes from linear to non-linear, and with an emphasis on these current sheet parameters, has yet to be performed. As we will show, the current sheet length is also an important parameter, and in this study we focus on the two parameters of magnetic shear and current sheet length, leaving current sheet thickness to a further study.

In this study, we perform high-fidelity numerical visco-resistive magnetohydrodynamic simulations
of the 3D tearing instability, covering the linear and non-linear phases, leading to highly dynamic magnetic reconnection and subsequent coronal heating. We perform a parameter study to investigate the effect of magnetic shear and current sheet length, and find that both are critical parameters for the onset of energy release due to non-linear 3D tearing. These results have important consequences for the nature of the onset of reconnection-driven heating and dynamics in the solar corona under various conditions.

\section{MHD Simulations}
\label{sec:num}
To study magnetic reconnection in the plasma of the solar corona in the fluid/MHD regime, we solve numerically the following visco-resistive MHD equations, written here in Lagrangian form:

\begin{eqnarray}
\frac{D\rho}{Dt} & = & -\rho\nabla.\mathbf{V}, \\
\frac{D\mathbf{V}}{Dt} & = & -\frac{1}{\rho}\left[\nabla P
+ \mathbf{J}\times\mathbf{B} + \mb{F}_{shock}\right],\\
\frac{D\mathbf{B}}{Dt} & = & (\mathbf{B}.\nabla)\mathbf{V}
- \mathbf{B}(\nabla .\mathbf{V}) - \nabla \times (\eta\mathbf{j}), \\
\frac{D\epsilon}{Dt} & = & \frac{1}{\rho}\left[-P\nabla .\mathbf{V}
 + H_{visc} \right],
\label{eqn:energy_MHD}
\end{eqnarray}
where $\rho$ is the mass density, $\mathbf{v}$ the velocity, $\mathbf{B}$ the magnetic field, and $\epsilon$ the internal specific energy density. The current density is given by $\mathbf{J}=\nabla\times\mathbf{B}/\mu_{0}$, where $\mu_{0}$ is the permeability of free space, and $\eta$ is the resistivity. The gas pressure, $P$, and the specific internal energy density, $\epsilon$, can be written as
\begin{eqnarray}
P & = & \rho k_{B}T/\mu_{m}, \\
\epsilon & =  & \frac{k_{B}T}{\mu_{m}(\gamma-1)}
\label{eqn:eos}
\end{eqnarray}
 respectively, where $k_{B}$ is Boltzmann's constant and  $\gamma$ is 5/3. The reduced mass $\mu_{m}$, for this fully ionized Hydrogen plasma, is given by $\mu_{m}=m_{p}/2$ where $m_{p}$ is the mass of a proton. {\color{black} The reader should note that the standard Omic heating term $\eta J^2$ has been removed from Eqn. (\ref{eqn:energy_MHD}), which is discussed below. }

The equations are solved using a Lagrangian-Remap (LaRe) approach \citep{Arber_2001}. 
The shock viscosity $\mathbf{F}_{shock}$ in the momentum equation is finite at discontinuities but zero for smooth flows, and the shock jump conditions are satisfied with an appropriate choice of shock viscosity \citep{Caramana_1998,Arber_2001}. This allows heating due to shocks to be captured as a viscous heating $H_{visc}$. This approach has been used in simulations of wave heating in the chromosphere \citep{Brady_2016}. 

The equations above are non-dimensionalized by dividing each variable ($C$) by its normalizing value ($C_{0}$). The set of equations requires a choice of three normalizing values. We choose normalizing values for the length, $L_{0}=10^{6}~ \textrm{m}$,  density, $\rho_{0}=1.5\times10^{-13} ~ \textrm{kg}/\textrm{m}^{3}$, and magnetic field, $B_{0}=10^{-3} ~ \textrm{T}$ ($10 ~ \textrm{G})$. This leads to a normalizing velocity of $V_{0} = 2.3\times10^{6} \textrm{m}/\textrm{s}$. However, all results presented in this study will use dimensional units.

To allow for rigorous validation and analysis of the numerical simulations, we use a simple initial equilibrium. The background density and temperature are $1.67\times10^{-13} ~ \textrm{kg}/\textrm{m}^{3}$ and $10^{6} ~ \textrm{K}$. The magnetic field is a modified 1-D, force-free Harris current sheet with a guide field,
\begin{eqnarray}
B_{x}(y) & = & {B_{x,0}}\tanh{(\frac{y}{a})}\cos{(\pi \frac{y}{L_{y}/2})}, ~ \textrm{for} ~ |y| < L_{y}/4\\
B_{z}(y) & = & \sqrt{(B_{z,0}^{2}-B_{x}(y)^{2})}
\end{eqnarray}
The cosine dependence on y reduces the shear component of the magetic field ($B_{x}$) to zero at $\pm L_{y}/4$ and allows for the use of periodic boundary conditions in $y$ as well as the other two dimensions. $B_{x,0}$ and $B_{z,0}$ are the values far from the center of the sheet. They define a shear angle $\theta = 2 arctan(B_{x,0}/B_{z,0})$, which is the amount that the magnetic field rotates across the sheet.


The simulation grid extends [-50,50] Mm in the $z$ direction (along the guide field), [-50,50] Mm in the $y$ direction (across the current sheet), and [-$L_{x}/2$,$L_{x}/2$] Mm in the $x$ direction (along the reconnecting field). As part of our study, we vary $L_{x}$, as, alongside the magnetic shear, the ratio of the most dominant wavelength of the tearing instability to this length turns out to be a critical parameter.

The $y$ domain has a non-constant resolution, where the grid size is a cubic function of the $y$ index. The minimum cell size in $y$ is 0.0125 Mm and the maximum is 0.34 Mm. The initial current sheet of half width $a=0.5 ~ \textrm{Mm}$ is well resolved as there are 75 cells across the full width in the $y$ direction.

Table \ref{tab:sims} shows the small parameter study performed. We vary both the length of the current sheet and the initial shear in the 3D magnetic field. As will be shown this leads to simulations with vastly different evolutionary paths.

\begin{table}[h]
    \centering
    \begin{tabular}{|c|c|c|c|c|}
    \hline
     Name & $L_{x}$ (Mm) & $B_{x,0}$ (G) & $B_{z,0}$ (G) & $\theta$ \\
    \hline
    \hline
    1   &  100 & 7.0 & 10.6  & 80 \\
    2   &  100 & 3.6 & 10.6  & 40 \\
    3   &  100 & 1.9 & 10.6  & 20 \\
    \hline
    4   &  20 & 7.0 & 10.6 & 80 \\
    5   &  20 & 3.6 & 10.6  & 40 \\
    6   &  20 & 1.9 & 10.6  & 20 \\    \hline
    \end{tabular}
    \caption{Pertinent simulations}
    \label{tab:sims}
\end{table}

Given that the resistivity of the solar corona is so low, our numerical scheme will have a numerical resistivity much larger than the Sun's actual resistivity. Because we wish to study physical effects that depend on the resistivity, we must use an explicit value that is larger than the numerical value. We adopt $\eta=1.24\times10^{2} ~ \Omega.\textrm{m}$, or a diffusivity of $D=\eta/\mu_{0}=10^{8}  ~ \textrm{m}^2 \textrm{s}^{-1}$, and have verified that the current sheet diffuses  at a rate consistent with this value. For a current sheet  half width ($a$) of $5\times10^{5} ~ \textrm{m}$ the diffusion time is $t_{d}=a^{2}/D = 2500 ~ \textrm{s}$, which, as will be shown, is longer than the tearing growth time, but still of the order of the duration of our simulation. Previous studies have mitigated this diffusion by removing the background diffusion of the original current sheet by inserting an additional electric field into the magnetic field evolution equation \citep[e.g.,][]{Del_Zanna_2016}. We do not perform such a  modification,  and so consistently follow the evolution of the slowly diffusing current sheet as it becomes unstable.

{\color{black} For the chosen value of $D= \eta/\mu_{0}=10^{8} ~ \textrm{m}^2 \textrm{s}^{-1}$, and for the parameter realization of $L=100 ~ \textrm{Mm}$ and  $B_{x,0}=7.0 ~ \textrm{G}$, and defining the Alfv\'{e}n velocity in terms of the shear field, $V_{a} = B_{x,0}/\sqrt{\rho_0\mu_{0}}$, a typical Lundquist number in our simulations is $S=L V_{a}/D$ is $S=1.52\times10^{6}$. We note that by design, our current sheet is not undergoing steady Sweet-Parker reconnection, but rather is in an initial static equilibrium that is unstable to the tearing mode. For a discussion of the onset of tearing instability in an evolving 2D system which has a pre-existing steady reconnection pattern, see \citet{Huang_2017}. }



These choices of physical and closure parameters lead to a non-negligible Ohmic dissipation of magnetic energy, whereas in the solar corona it is negligible on all but very small scales. Therefore, while we lose energy from the magnetic field via Ohmic diffusion, {\color{black} we have removed the standard Ohmic $\eta J^{2}$ heating term from Eqn. \ref{eqn:energy_MHD} for the internal energy density $\epsilon$}.
As a consequence, the only form of heating in our simulation is caused by the shock heating described above. Thus the energy flow during magnetic reconnection is from magnetic energy into bulk kinetic energy and then into thermal energy via viscous heating at small spatial scales.  The tearing mode will create multiple islands of reconnecting flux, and the interaction of these plasmoids and associated flows will generate gradients in velocity leading to plasma heating. 

{\color{black} In order to assess the fidelity of the simulations we have calculated the numerical energy lost in the simulations. In resistive MHD, for a triply periodic system with no gravitational force and no other themodynamic terms (such as radiation), the total energy $E=\frac{\rho V^2}{2} + \rho\epsilon + \frac{B^2}{2\mu_{0}}$  is conserved: $\frac{\partial E}{\partial t} = 0$ and any numerical losses can be estimated by $E(t)-E(0)$. In our modified system where the Ohmic heating is not included in the energy equation then $\frac{\partial E}{\partial t} = -\eta J^2$. Therefore we can estimate the energy numerical losses by $E(t)-E(0) + \int_{0}^{t}{\eta J(t')^2 dt'}$. } We have performed checks to ensure that this numerical loss is less than the smallest physical energy (in this case the kinetic energy is typically smaller than both the magnetic and internal energy), so that our system evolution is considered resolved and consistent with the equations solved.

To initiate the linear stage of the 3D tearing instability, the velocity $V_{y}$ is set to be a superposition of weak Fourier modes, with mode numbers $(m,l,n)$,
which encompass the predicted allowed dominant parallel and oblique modes in this particular system, based on the linear theory below \citep{Baalrud_2012}:
 \begin{eqnarray}
 V_{y}(x,y,z) & = & \sum_{m,l,n}{V_{0}}^{mln} \sin{(k_{x}x-{\phi_{x}}^{mln})}\\ 
  & & \sin{(k_{y}y-{\phi_{y}}^{mln})}\sin{(k_{z}z-{\phi_{z}}^{mln})}
 \end{eqnarray}
where $k_{x} = 2\pi/\lambda_{x}=2\pi m/L_{x} ~( m=[1,n_m])$, $k_{y} = 2\pi/\lambda_{y} = 2\pi l /L_{y}~( l=[1,n_l])$, and $k_{z}=2\pi\lambda_{z} = 2\pi n/L_{z} ~(n=[0,n_n])$. To initiate a large range of modes we use $(n_m,n_l,n_n)=(25,5,5)$. The phases ($\phi^{mln}$) and amplitudes ($V_{0}^{mln}$) for each mode are chosen randomly, with the amplitudes chosen randomly from a top-hat function in the range [-10, 10] m~s$^{-1}$ This results in an average initial velocity of 43 m~s$^{-1}$.

Despite the use of this specific initial condition on $V_{y}$ we have performed tests with different values of $n_{n,m,l}$ and also using randomly seeded values of $V_{y}$, and find that the evolution of the system is very similar, as the fastest growing modes allowed by the system quickly dominate the dynamics mostly  independent of the initial $V_{y}$ distribution. 


\section{Linear Theory and Validation}
\label{sec:validate}
\begin{figure}
\begin{center}
\includegraphics[width=0.20\textwidth]{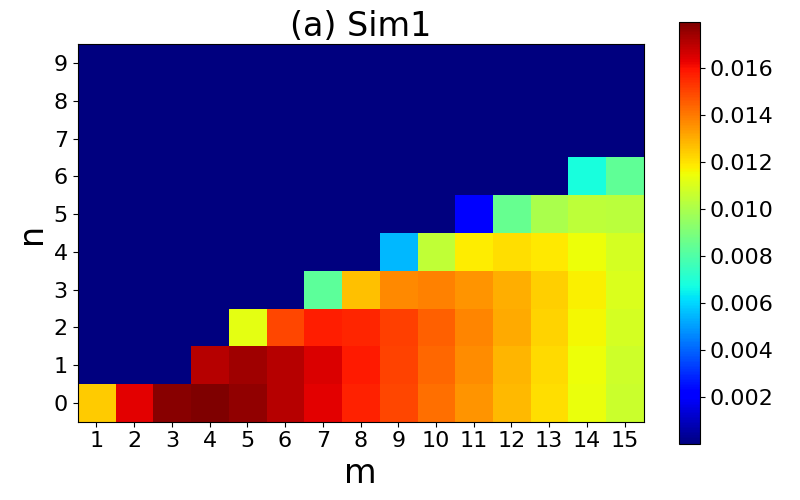}
\includegraphics[width=0.20\textwidth]{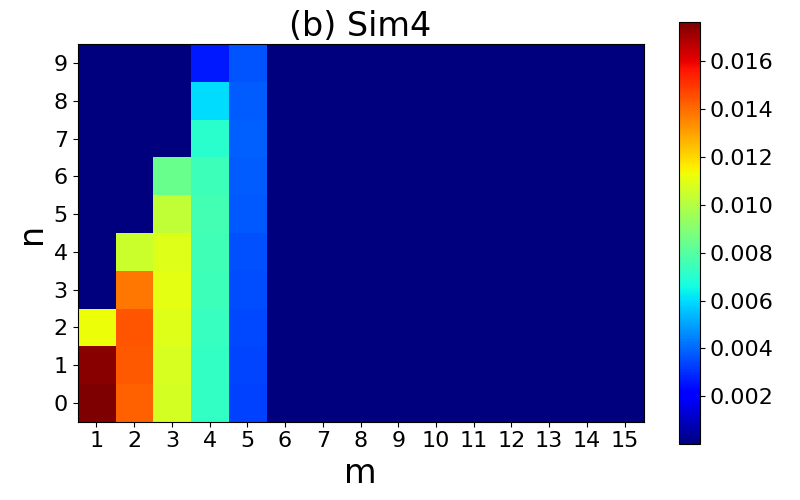}\\
\includegraphics[width=0.20\textwidth]{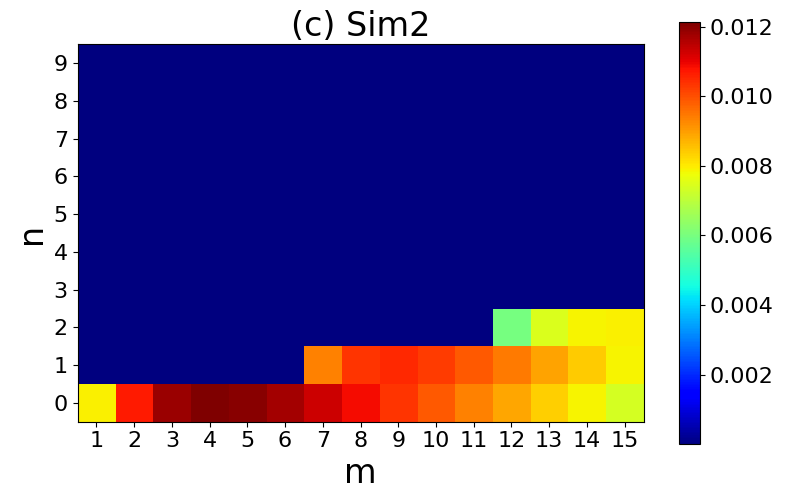}
\includegraphics[width=0.20\textwidth]{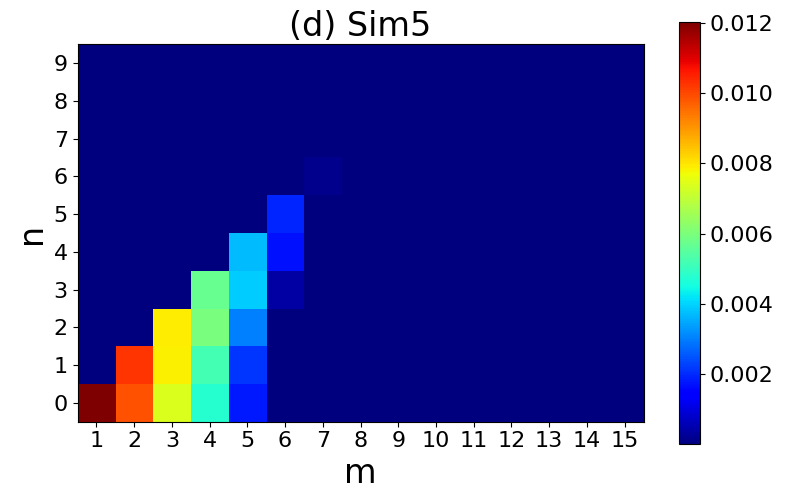}\\
\includegraphics[width=0.20\textwidth]{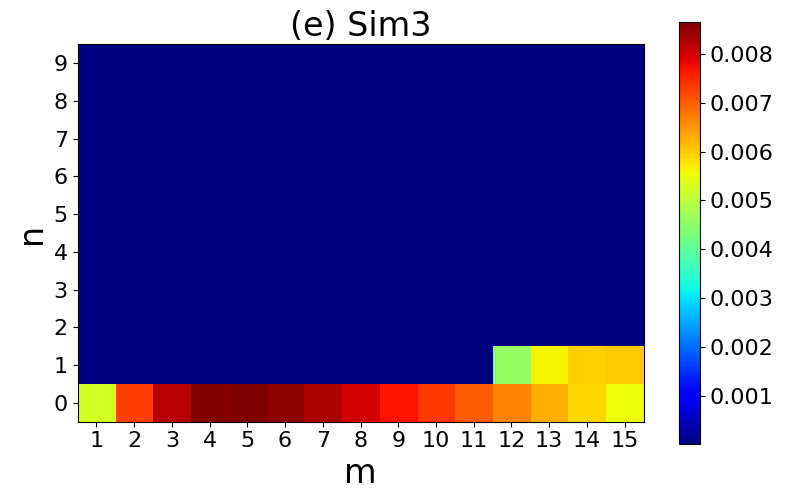}
\includegraphics[width=0.20\textwidth]{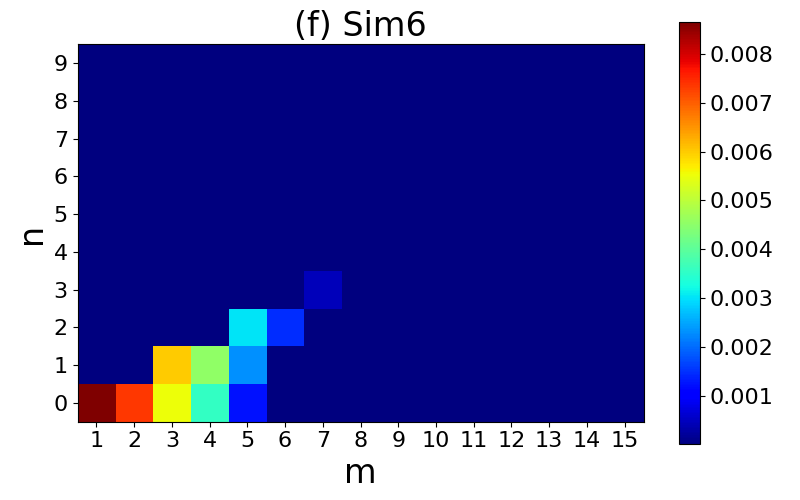}
\caption{Predicted growth rates of various ($m,n$) modes in Simulations 1-6. Simulations 1-3 are on the left, 4-6 are on the right. The growth rate is given in $s^{-1}$. \label{fig:theory_rates}}
\end{center}
\end{figure}

Before discussing the non-linear evolution of the six simulations, it is critical to validate the linear evolution of the simulations against theoretical predictions.

\begin{figure}
\begin{center}
\includegraphics[width=0.4\textwidth]{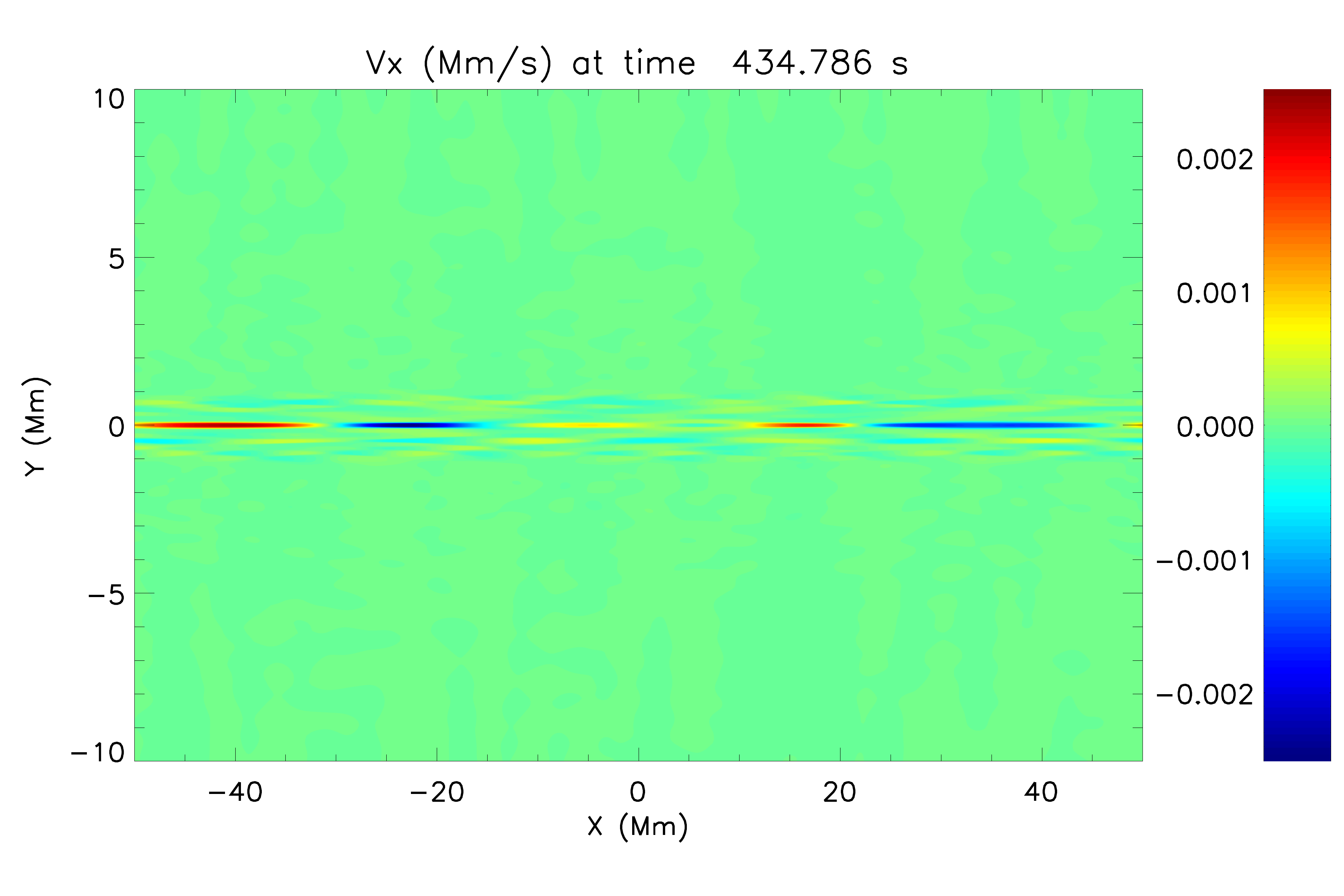}
\includegraphics[width=0.4\textwidth]{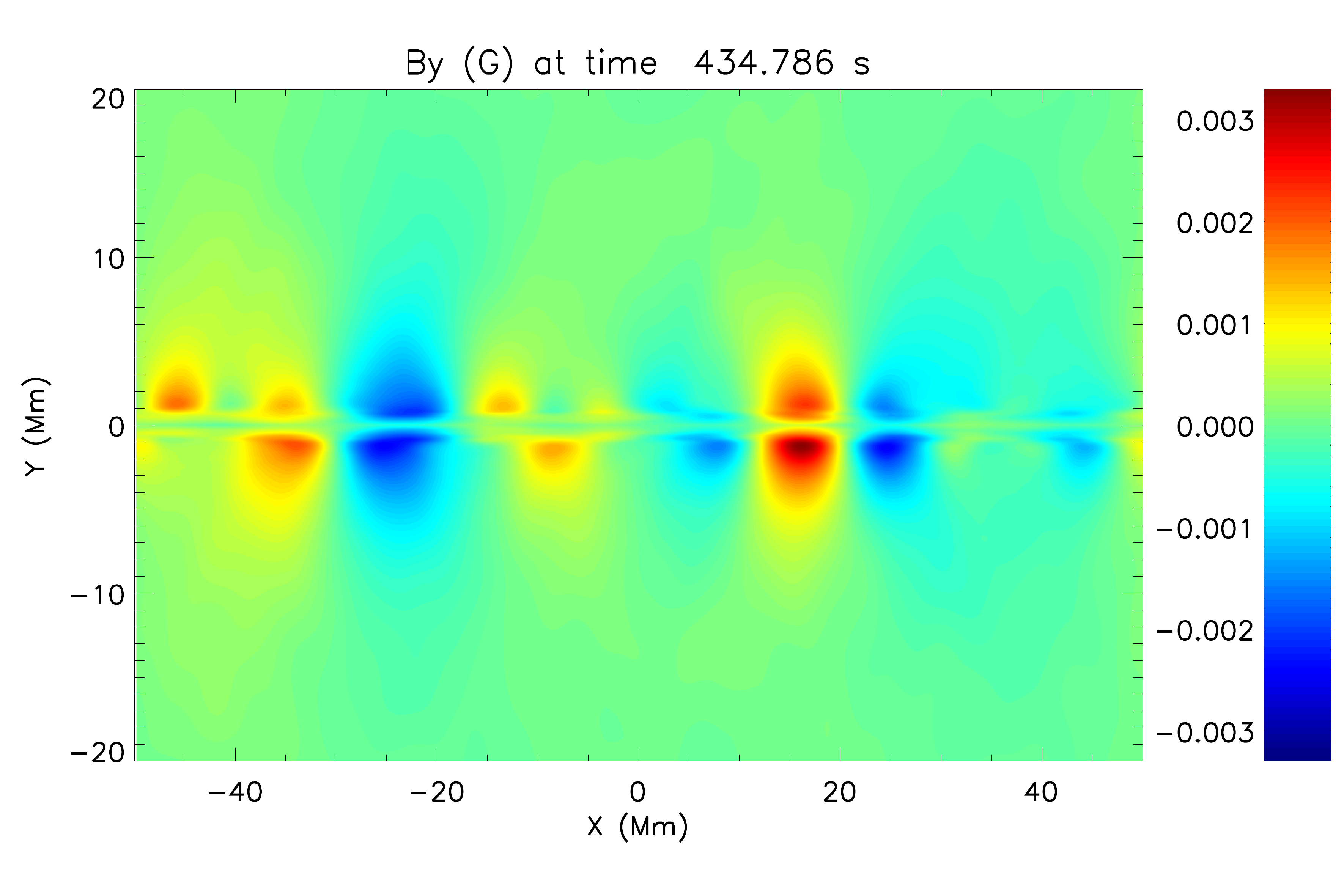} \\
\caption{Snapshot of $V_{x} (Mm/s)$ and $B_{y} (G)$ in the linear stages of the tearing instability for Simulation 1. The 2D slice is taken at $z=0$.  \label{fig:2D_early_evolution}}
\end{center}
\end{figure}

 As is well known, the tearing instability forms a chain of magnetic islands, which are the cross-sections of flux tubes in 3D. Each tearing mode consists of identical tubes that are lined up side by side within a plane. The plane is parallel to the current sheet and can be located at the center of the sheet or offset from it. The tubes are aligned with the local magnetic field direction, and since the field rotates across the sheet, the tubes have a variable orientation. Modes at the center of the sheet ($y = 0$) are known as parallel modes because the tubes are parallel to the guide field ($B_z$), while modes offset from the center are known as oblique modes. Each mode is characterized by wavenumbers $k_x$ and $k_z$, which are the inverse wavelengths along the sheet and along the guide field, respectively:  $k_x = 2\pi/\lambda_x, k_z = 2\pi/\lambda_z$. The tilt of the tubes relative to the guide field direction is given by $\arctan(k_z/k_x) = \arctan(B_x/B_z)$. It is convenient to identify the modes by a pair of mode numbers $(m, n)$, which indicate the number of wavelengths that fit along the dimensions $L_x$ and $L_z$, respectively:  $m = L_x/\lambda_x = L_x k_x/(2\pi), n = L_z/\lambda_z=L_z k_z/(2\pi)$. Parallel modes have $n = 0$, while oblique modes have $|n| > 0$. Henceforth, we will only refer to positive n, but the negative counterpart located on the other side of the sheet is implied. Note that the m and n mode numbers are the same as those used to define the Fourier components of the applied $V_y$ perturbations.

 The reason for the different tearing modes and their properties can be understood as follows. Each island (tube) is formed by reconnection at a pair of X-points (X-lines). The extremely small resistivity is unimportant everywhere except near the X-points, where it is critical. A standard perturbation analysis of the linearized induction equation, Equation (3), reveals that this occurs where $\mb{k}\cdot\mb{B} = 0$, i.e., where $k_z/k_x = -B_x/B_z$. This is sometimes referred to as a resonance surface. Further discussion of parallel and oblique tearing modes can be found in \citet{Baalrud_2012} and \citet{Onofri_2004}.


 The growth rate of each unstable mode is a function of the mode numbers $(m,n)$, and hence wavenumbers $(k_{x},k_{z})$, the current sheet half width $a$, the amount of magnetic shear, and the Lundquist number, $S$. Figure \ref{fig:theory_rates} shows the predicted growth rates of various allowed modes of the 3D tearing instability for the $t=0$ condition of all 6 simulations, calculated using the boundary-layer matching approach of \citet{Baalrud_2012}. Note that later in this study, because there is a non-negligible rate of diffusion of the background current sheet, the current sheet parameters of width and field strengths change. We fit the current sheet width and strength to the observed $B_{x}$ and $B_{z}$ profiles, and recalculate the growth rates for each $(m,n)$ pairing with these fitted parameters to give the instantaneous theoretical rates.

One can see from Figure \ref{fig:theory_rates} that for the simulations with larger $L_{x}$ (Sims. 1,2,3, Figure \ref{fig:theory_rates} panels (a),(c),(e)), the most unstable mode is a parallel mode ($n=0$) with an $m$ value larger than 1 (typically 4 or 5). Hence there are multiple wavelengths of this dominant mode within the $x$ domain. In contrast, the most unstable parallel mode in the smaller $L_{x}$ simulations (Sims. 4,5,6, Figure \ref{fig:theory_rates} panels (b),(d),(f)) has $m=1$, indicating one wavelength in the $x$ domain. In addition, simulation 4, which has the strongest shear,  has a strongly unstable oblique mode $(m,n)=(1,1)$, but simulations 5 and 6, which have weaker shear, have weaker oblique modes. These facts will become extremely relevant when interpreting the overall evolution of these simulations.

 \begin{figure}
\begin{center}
\includegraphics[width=0.5\textwidth]{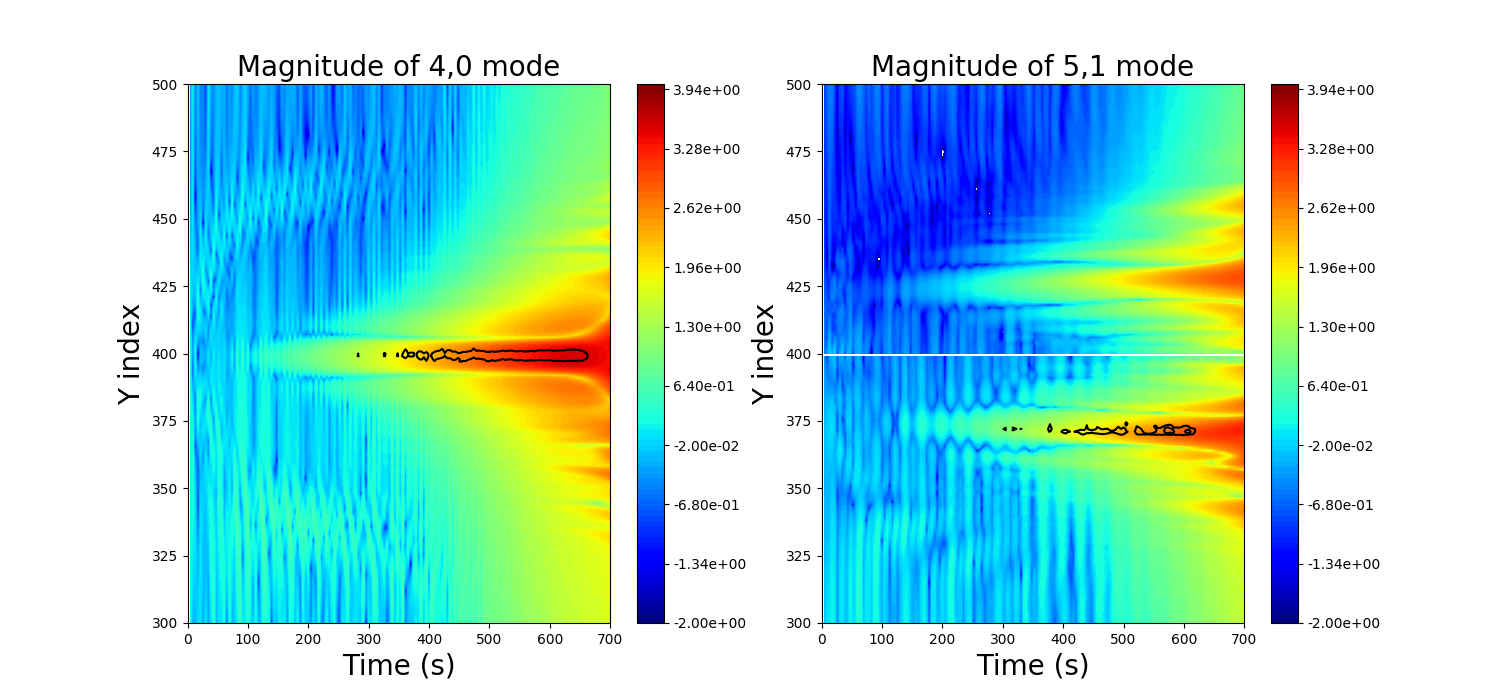}
\caption{Power in the (4,0) and (5,1) modes in Simulation 1, as a function of time and $y$-index.  The black contour is $\log_{10}\Gamma=2$, see Eqn. (\ref{eqn:gamma}). \label{fig:mode_locations}}
\end{center}
\end{figure}


Figure \ref{fig:2D_early_evolution} shows a snapshot of the early evolution of Simulation 1, in the linear stage of the instability. The images show $V_{x}$, which highlights the bi-directional outflows from the x-points between the tearing islands, and $B_{y}$, which is the reconnected field (field component across the sheet, created by reconnection), in the $z=0$ plane. The pattern of $V_{x}$ is strongest at $y=0$, as the strongest mode is a parallel mode, but one can also see a signal at $|y|>0$, which is primarily a result of the growth of an oblique mode. Although the theoretical dominant parallel mode is $m=4$ at this early stage for Simulation 1, the main signal for $V_{x}$ at $y=0$ appears to have 3 wavelengths. This is a result of the superposition of the most dominant (4,0) parallel mode with the sub-harmonic parallel modes which only have a slightly smaller growth rate, as well as the most dominant oblique mode. 


By taking Fourier transforms of the $V_{x}$ signal, one can decompose the signal into $(m,n)$ modes, and evaluate the power in each of these mode pairs as a function of $y$ and time.  The power in the dominant parallel (4,0) and oblique (5,1) modes in Simulation 1, is shown in Figure \ref{fig:mode_locations}. The power is shown as a function of time and the $y$-index of the nonuniform numerical grid (to zoom in on the center of the sheet). 
The locations of where these modes grow are locations of small $\mathbf{k}\cdot\mathbf{B}$. To visually validate that the modes are growing where the resonant surfaces are, a contour of a measure of the relative strength of the resistive term in the induction equation can be overlaid on these power plots. We therefore take Fourier transforms of the $z$ components of  $\mathbf{V}\times\mathbf{B}$ and $\eta\mathbf{J}$, calculate the power of these transforms for the relevant modes ( (4,0) or (5,1) ), and then calculate the normalized difference:
\begin{equation}
\Gamma = \frac{FFT((\eta\mathbf{J})_{z})-FFT((\mathbf{V}\times\mathbf{B})_{z})}{FFT((\mathbf{V}\times\mathbf{B})_{z})} \label{eqn:gamma}
\end{equation}
The black contour on each panel of Figure \ref{fig:mode_locations} is at a value of $\log_{10}(\Gamma) = 2$. As expected, it coincides with the peak of the power in each mode. Not shown is the contour for $y>0$.

\begin{figure}
\begin{center}
\includegraphics[width=0.5\textwidth]{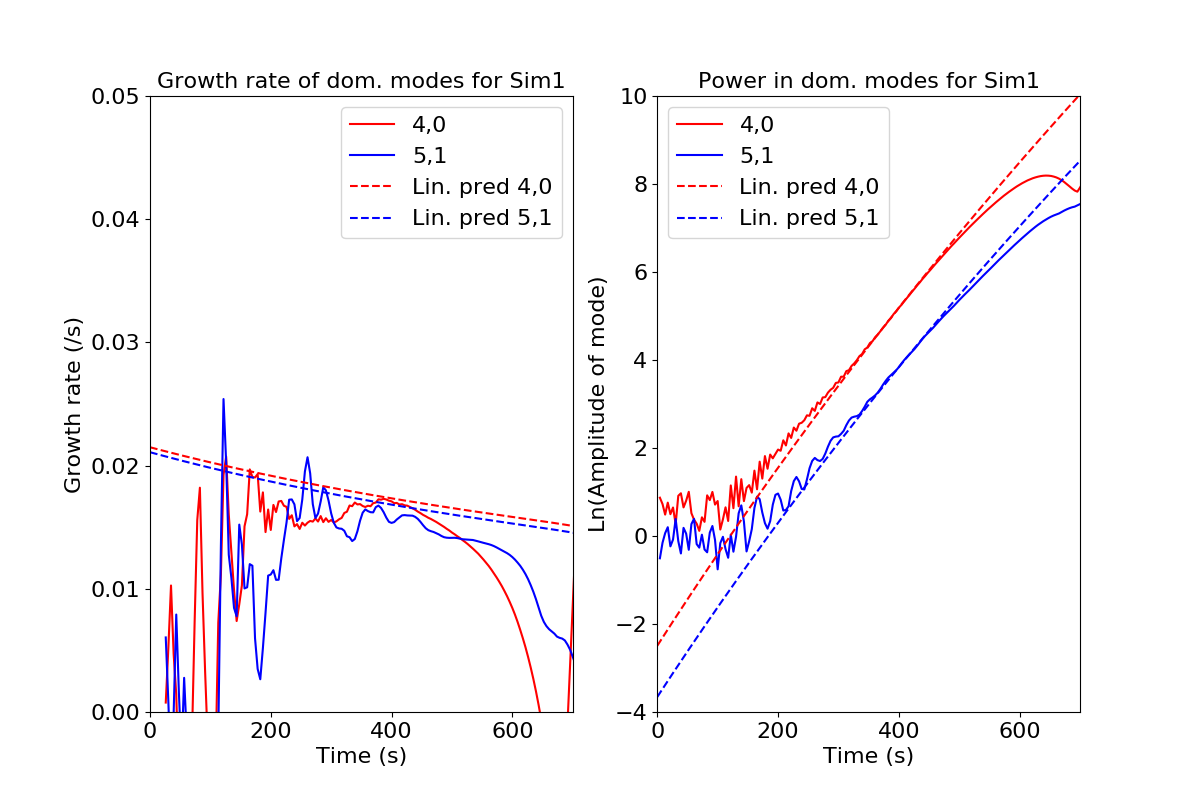}
\caption{Comparison of predicted growth rate and numerically simulated growth rate of the dominant modes in Simulation 1. \label{fig:validate}}
\end{center}
\end{figure}

We can now compare the growth of the modes in the simulations with the theoretical growth rates. From Figure \ref{fig:mode_locations} and from similar plots for the other modes, we take the maximum power over $y$ at each time to obtain a time series of the power in each mode, and then fit it locally to an exponential curve to derive a growth rate. This instantaneous growth rate is shown as solid lines in Figure \ref{fig:validate}. The theoretically derived linear growth rates for the modes are shown as dashed lines, and as discussed in \S \ref{sec:num}, take into account the changes in the growth rate due to the small but non-negligible diffusion of the equilibrium current sheet. Also shown is the power of the modes, with the theoretical value constructed from  the theoretical growth rates assuming that the simulation and theoretical curves match at t=400 s).

It is clear from Figure \ref{fig:validate} that there is a period around 0-200 s where the simulation is oscillating, as the equations process the initial condition random superposition of Fourier modes. 
Between 300 s and 500 s, the match between simulation growth rate and theoretical growth rate is good, before the system goes non-linear just after 400 seconds. This brief window of linear evolution between initial oscillations and non-linear evolution is a consequence of the fast growth time of the instability for the chosen parameters. For more slowly evolving systems, given by different choices of Lundquist number and sheet width, this window is larger. The generally good agreement between simulation and theory,  also seen in the other modes we examined, is a measure of confidence that the non-linear evolution in our numerical simulations is based on a realistic evolution of the linear stage. 

\section{Non-Linear Results}
\label{sec:results}

\begin{figure}
\begin{center}
\includegraphics[width=0.4\textwidth]{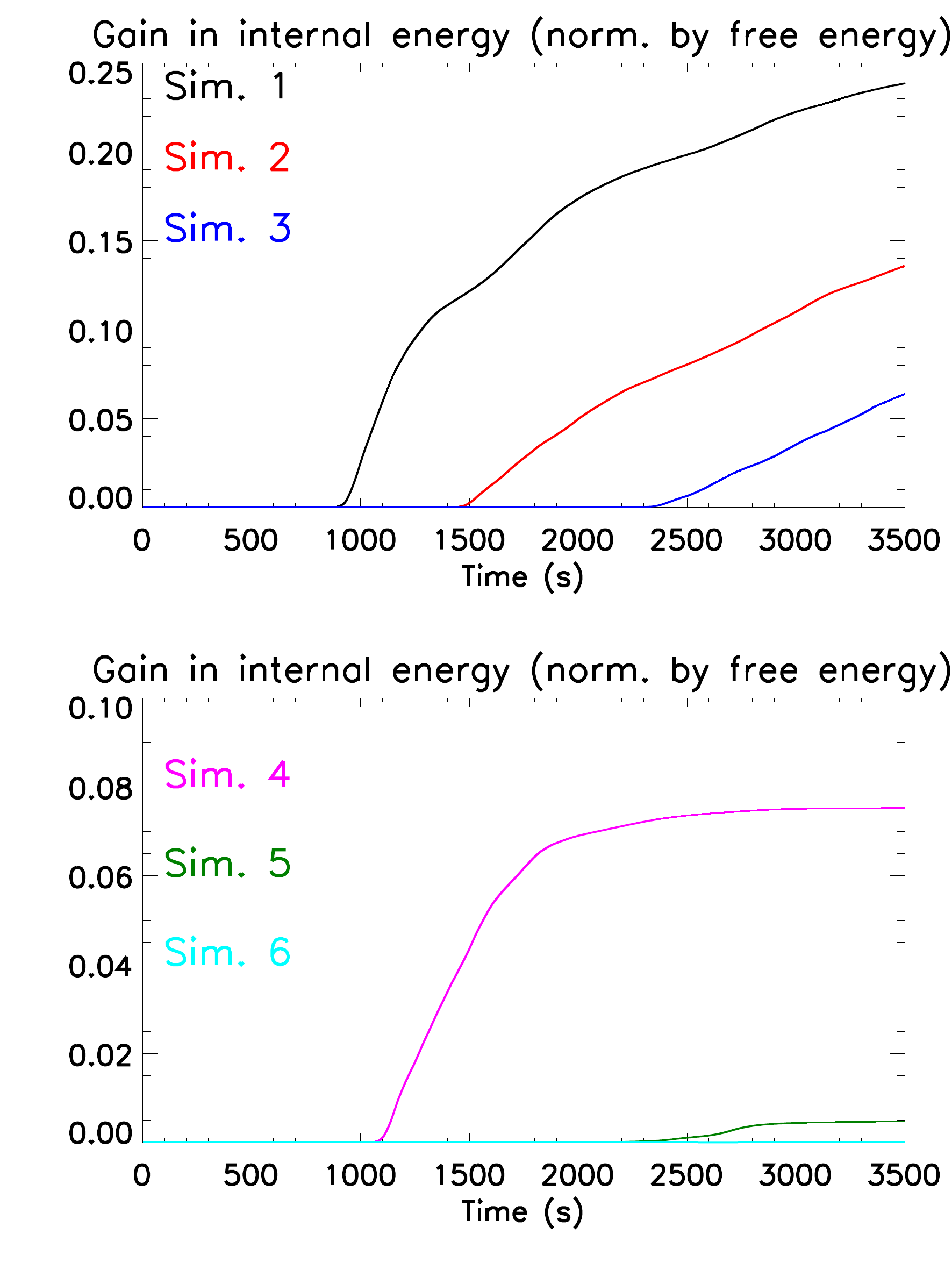}
\includegraphics[width=0.4\textwidth]{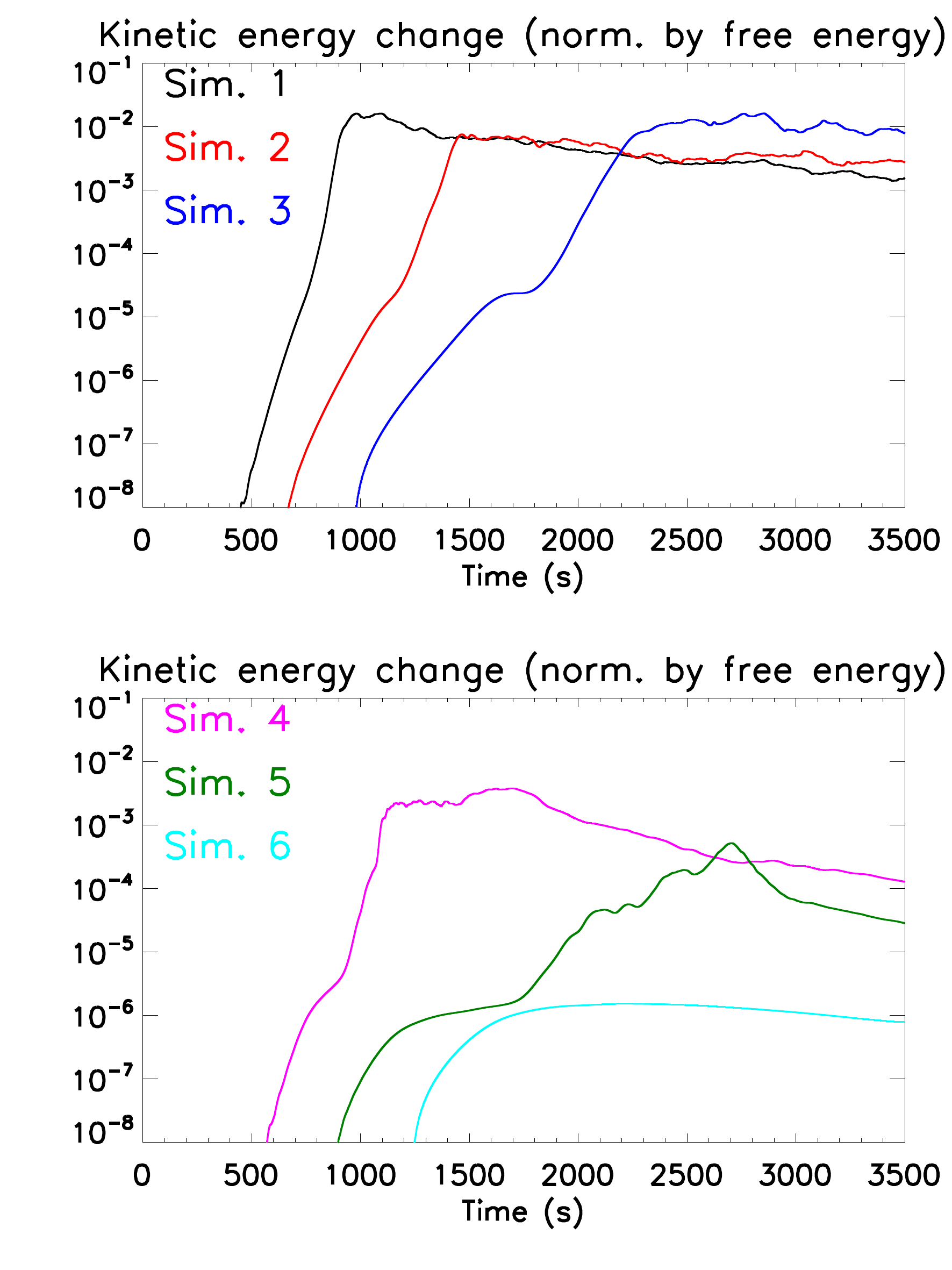}
\caption{Left: Gain in internal energy (heating) normalized by the initial free magnetic energy, for all six simulations. Right: Kinetic energy for the same six simulations. \label{fig:energy}}
\end{center}
\end{figure}

\begin{figure*}
\begin{center}
\includegraphics[width=0.4\textwidth]{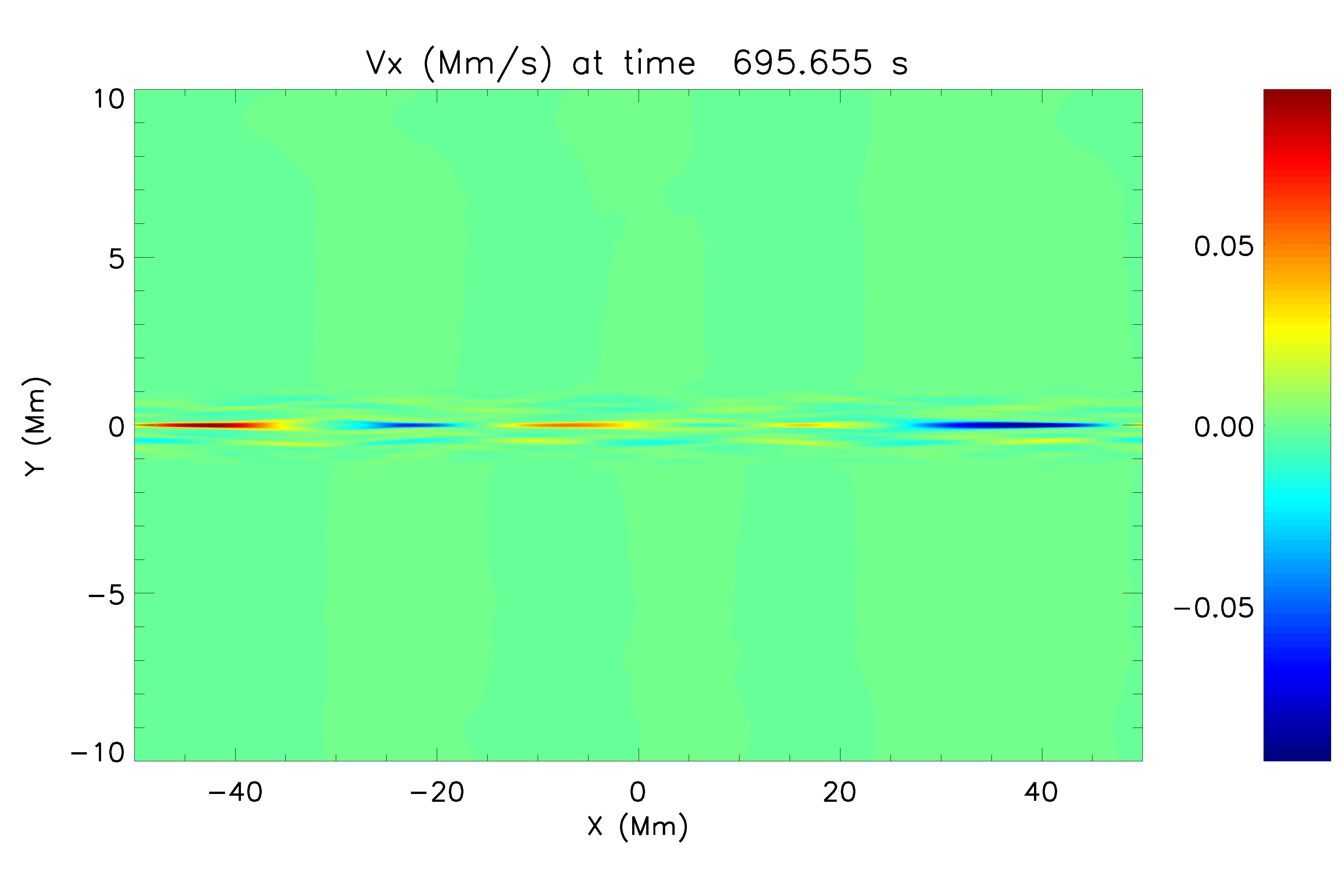}
\includegraphics[width=0.4\textwidth]{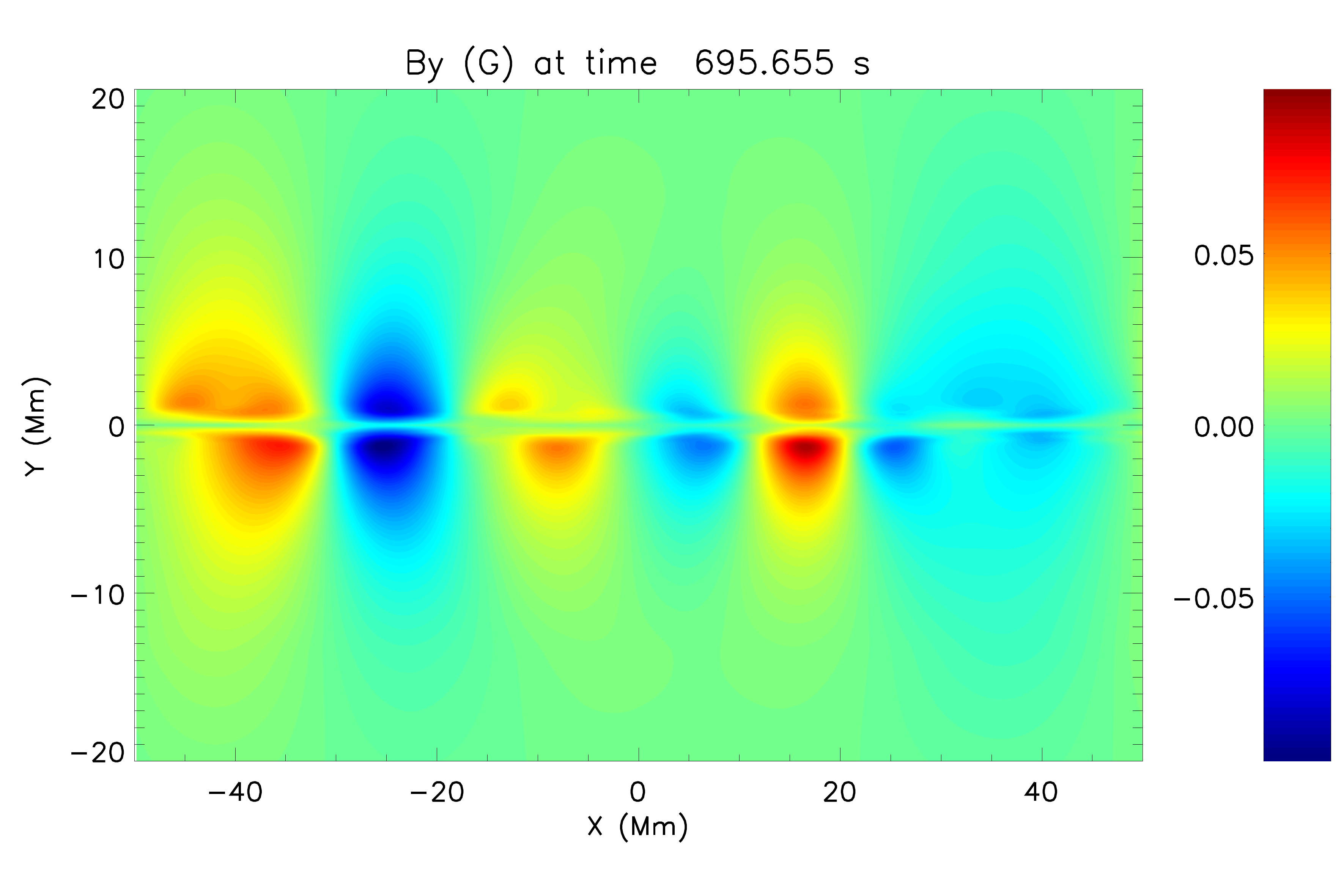} \\
\includegraphics[width=0.4\textwidth]{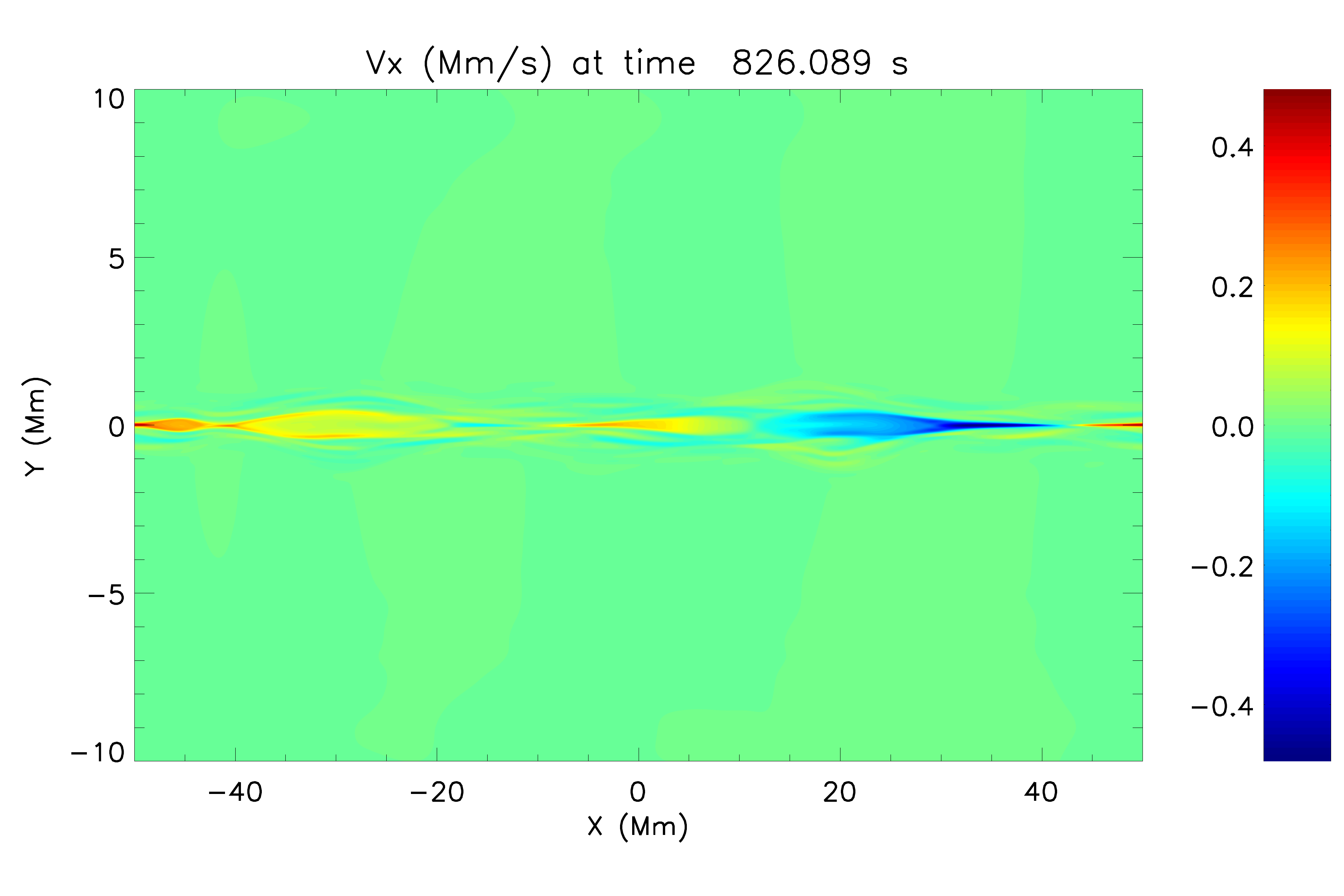}
\includegraphics[width=0.4\textwidth]{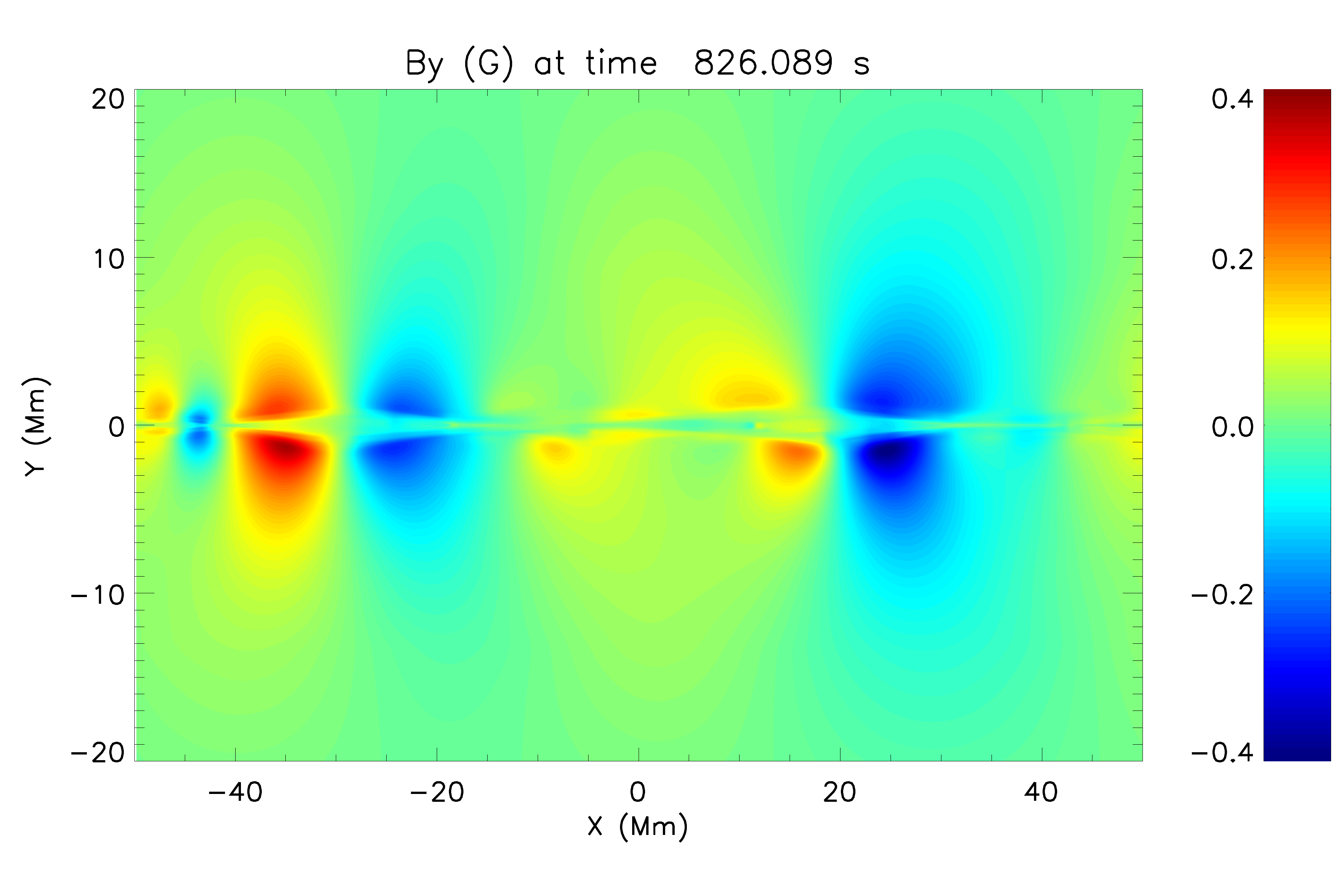} \\
\includegraphics[width=0.4\textwidth]{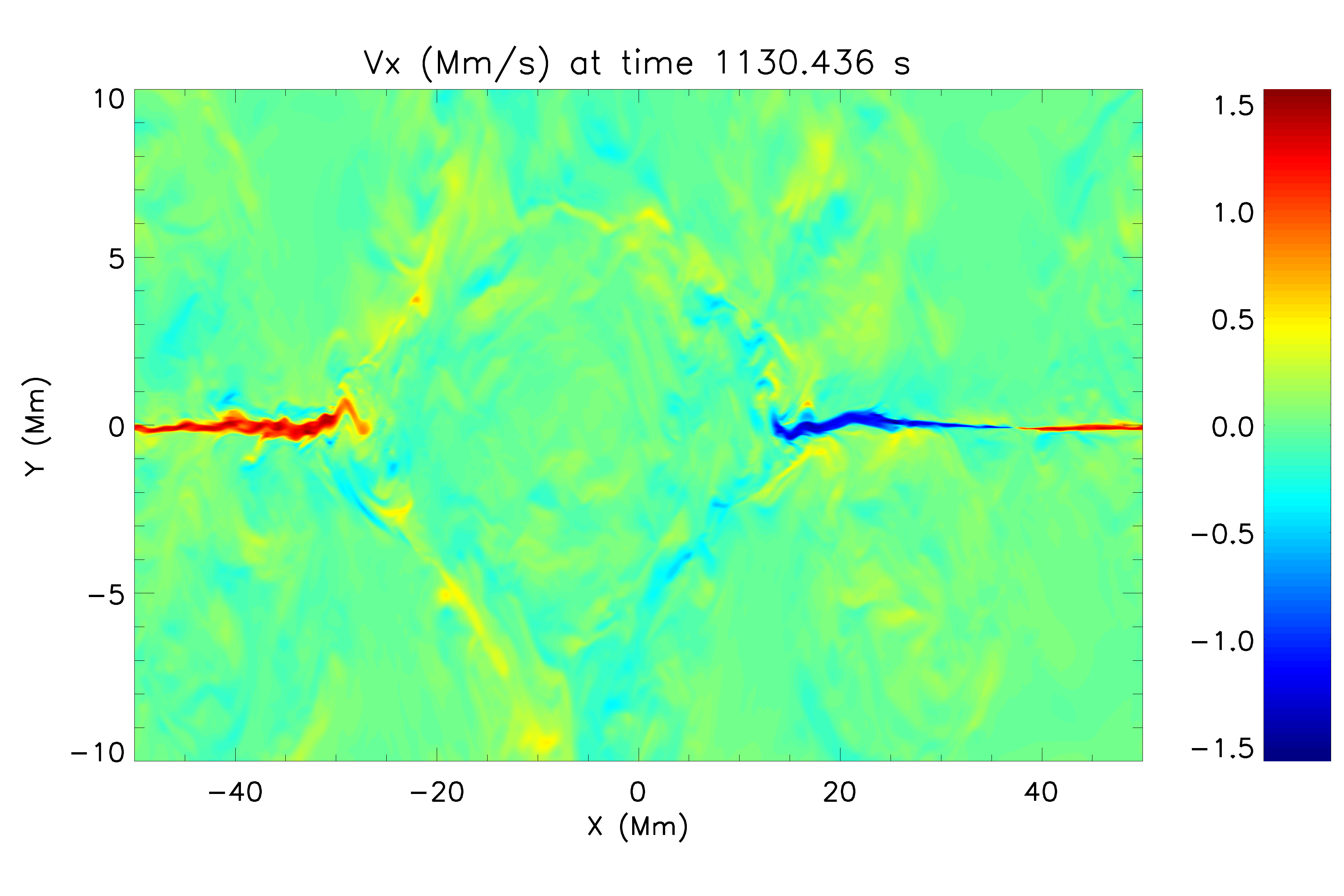}
\includegraphics[width=0.4\textwidth]{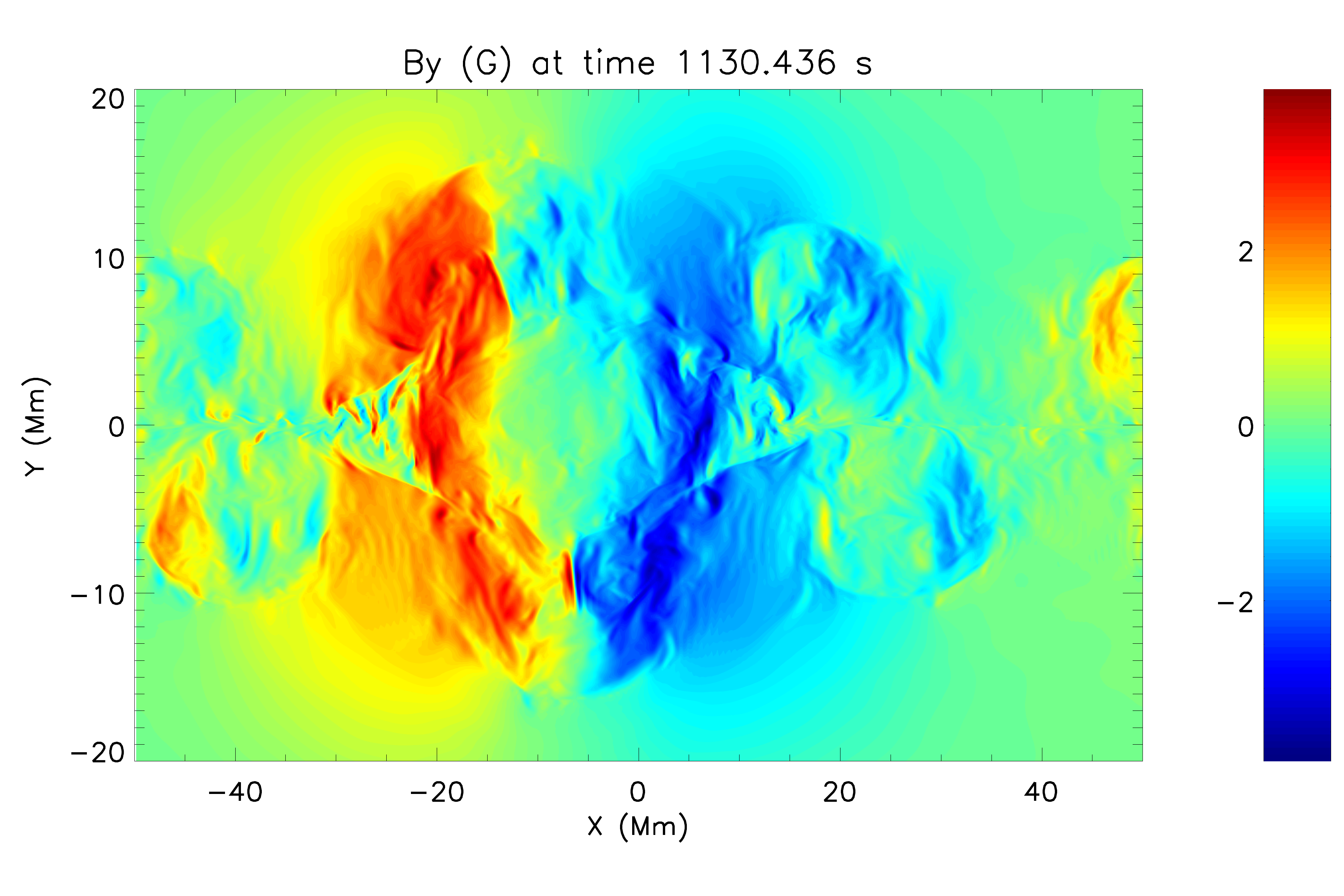} \\
\caption{$V_{x}$ and $B_{y}$ at $z=0$ for Simulation 1. \label{fig:highshear_2D}}
\end{center}
\end{figure*}

\begin{figure*}
\begin{center}
\includegraphics[width=0.4\textwidth]{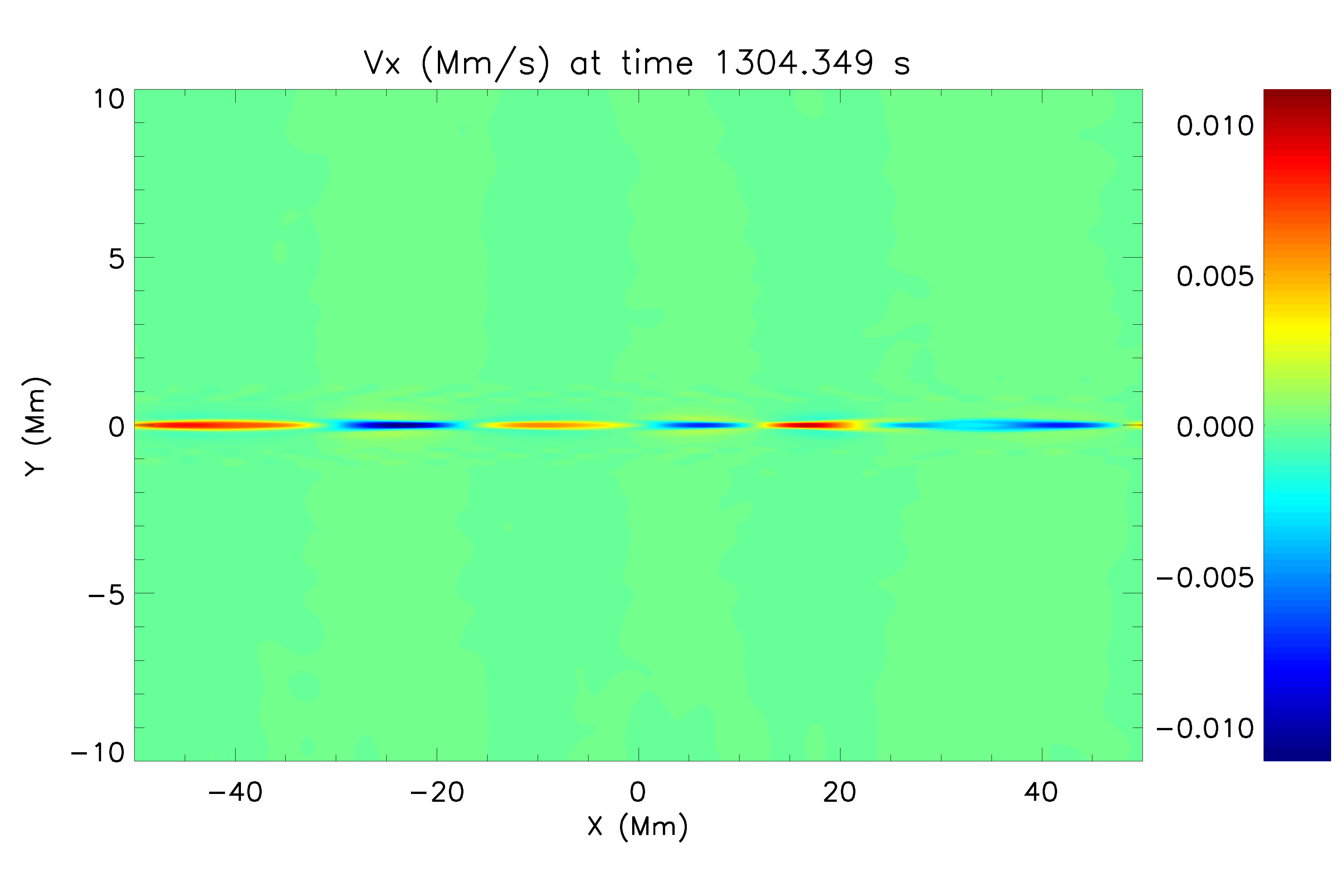}
\includegraphics[width=0.4\textwidth]{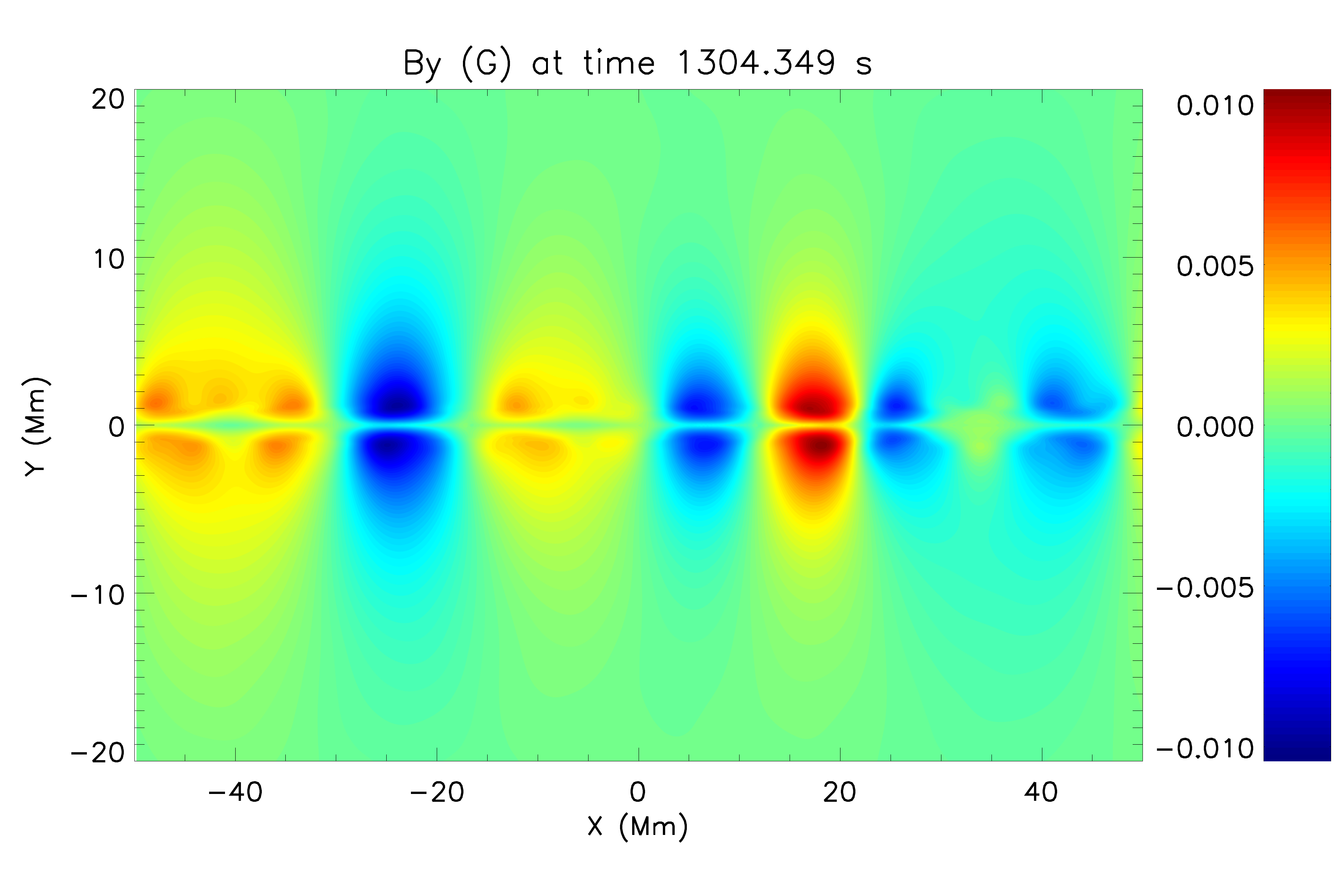} \\
\includegraphics[width=0.4\textwidth]{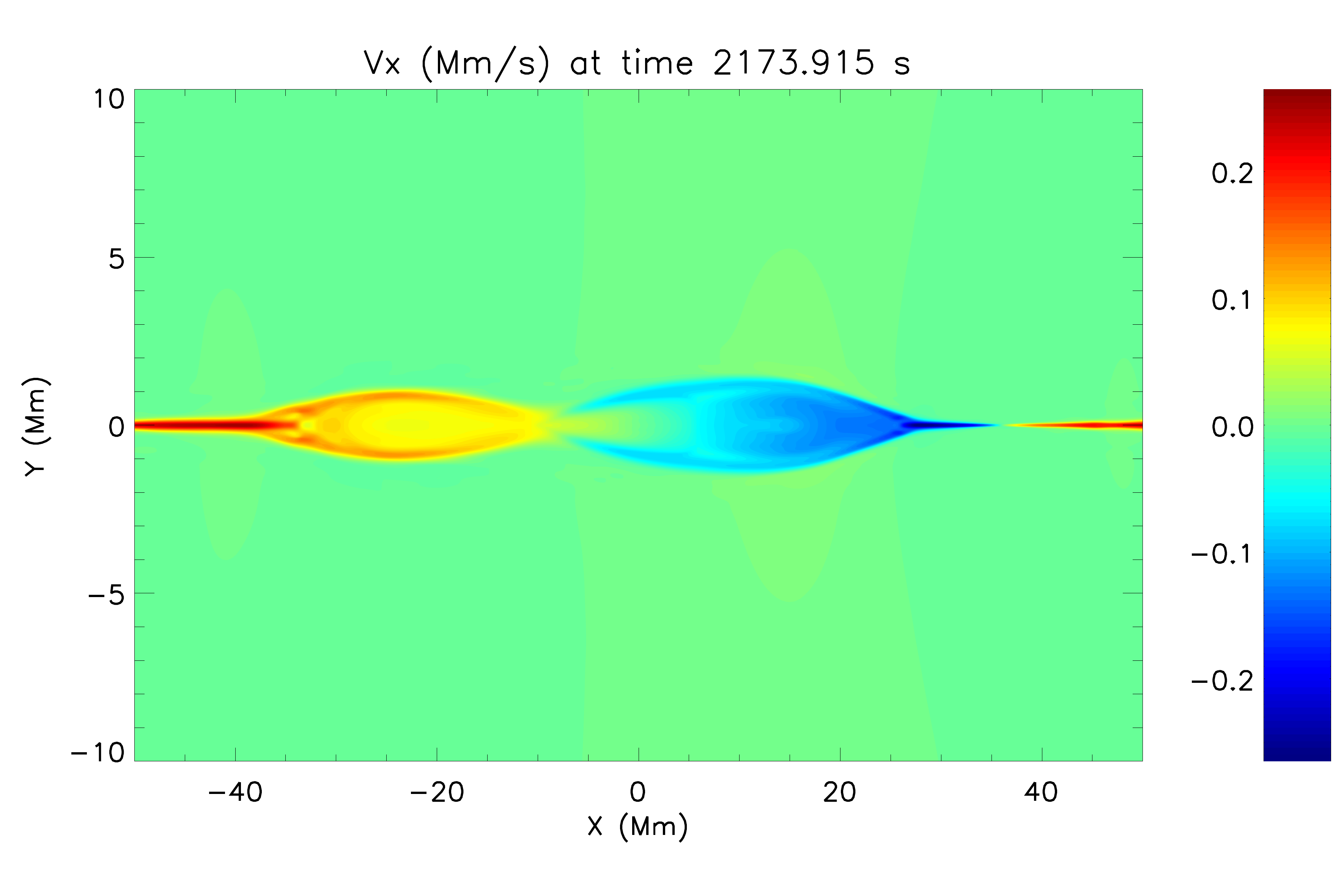}
\includegraphics[width=0.4\textwidth]{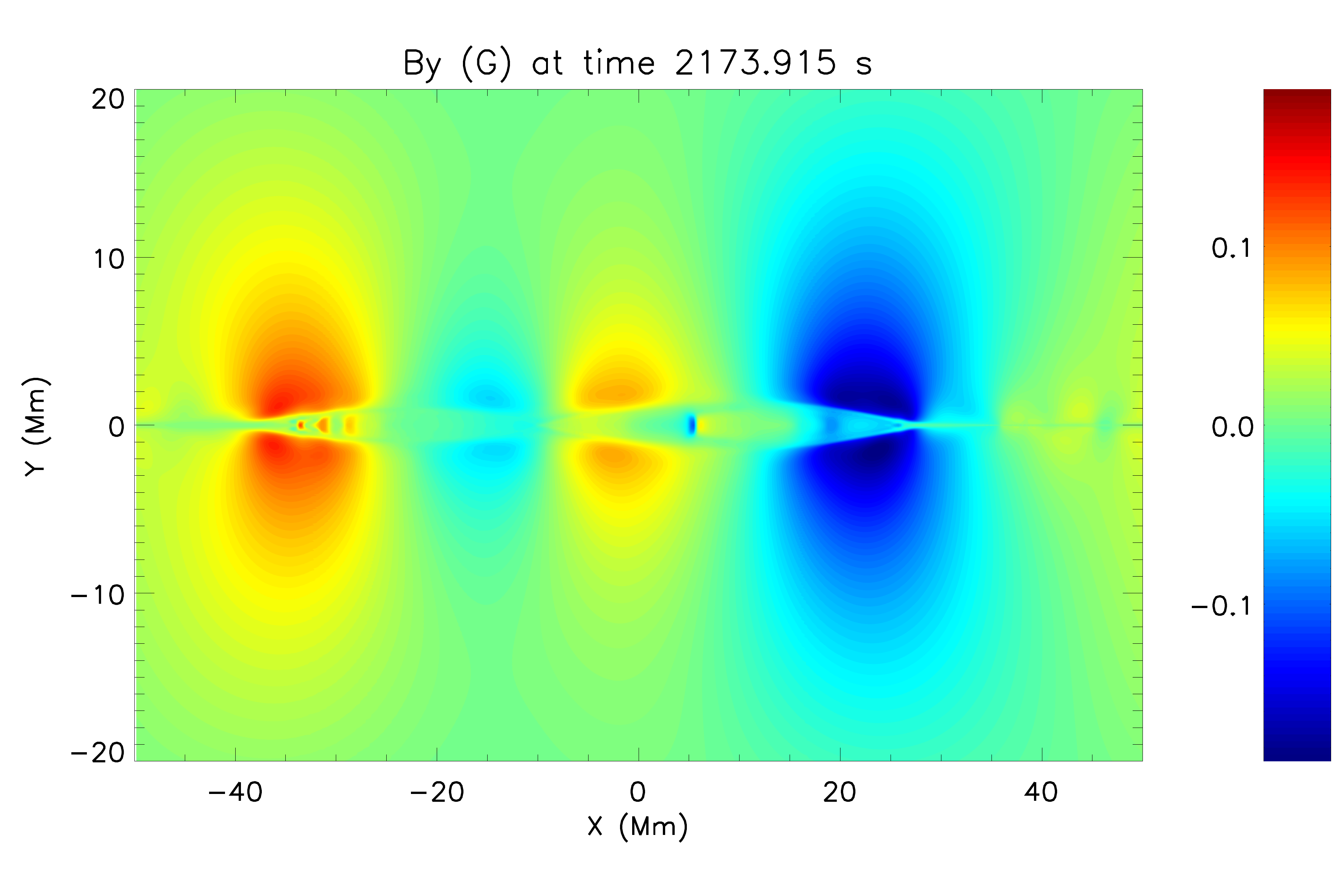} \\
\includegraphics[width=0.4\textwidth]{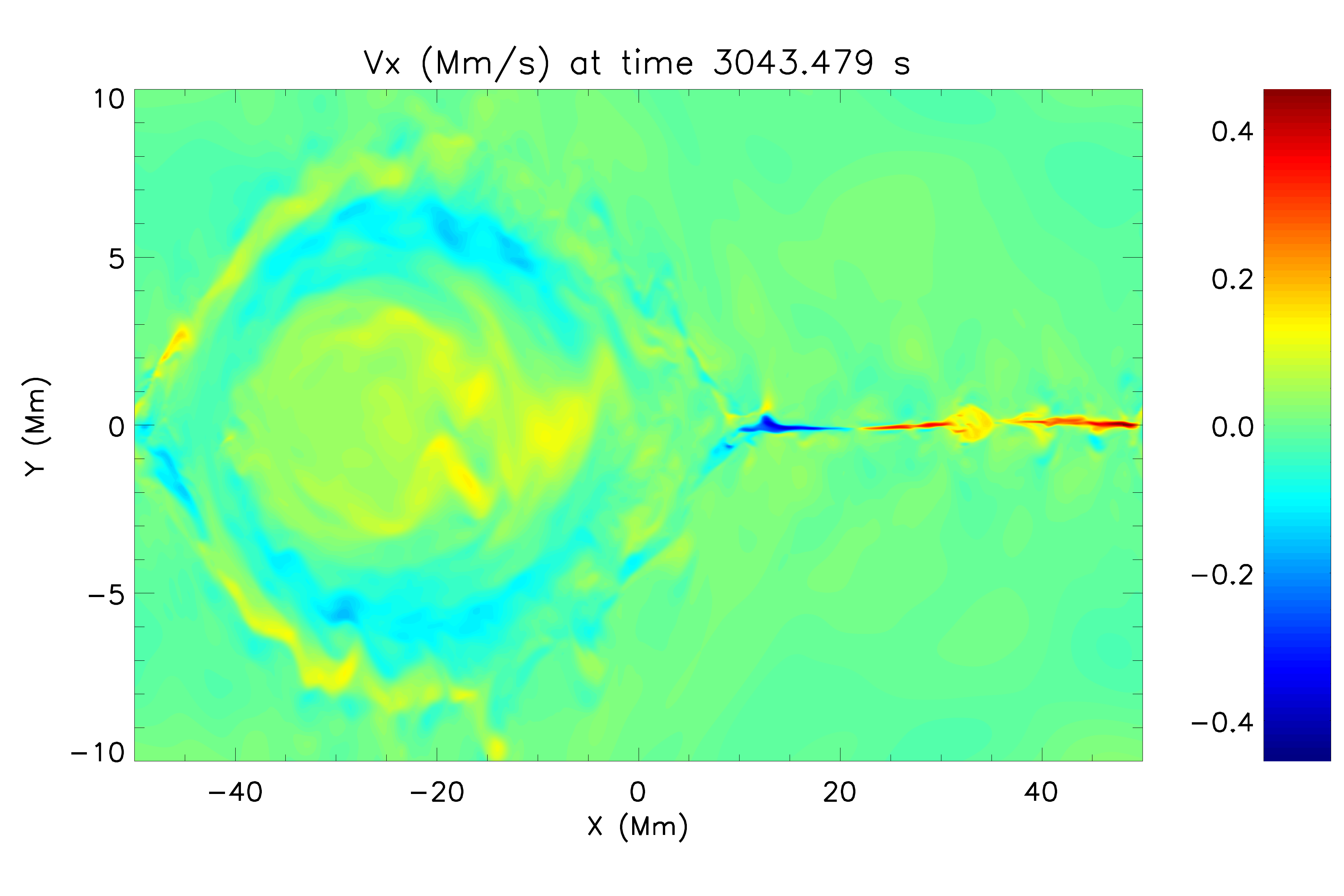}
\includegraphics[width=0.4\textwidth]{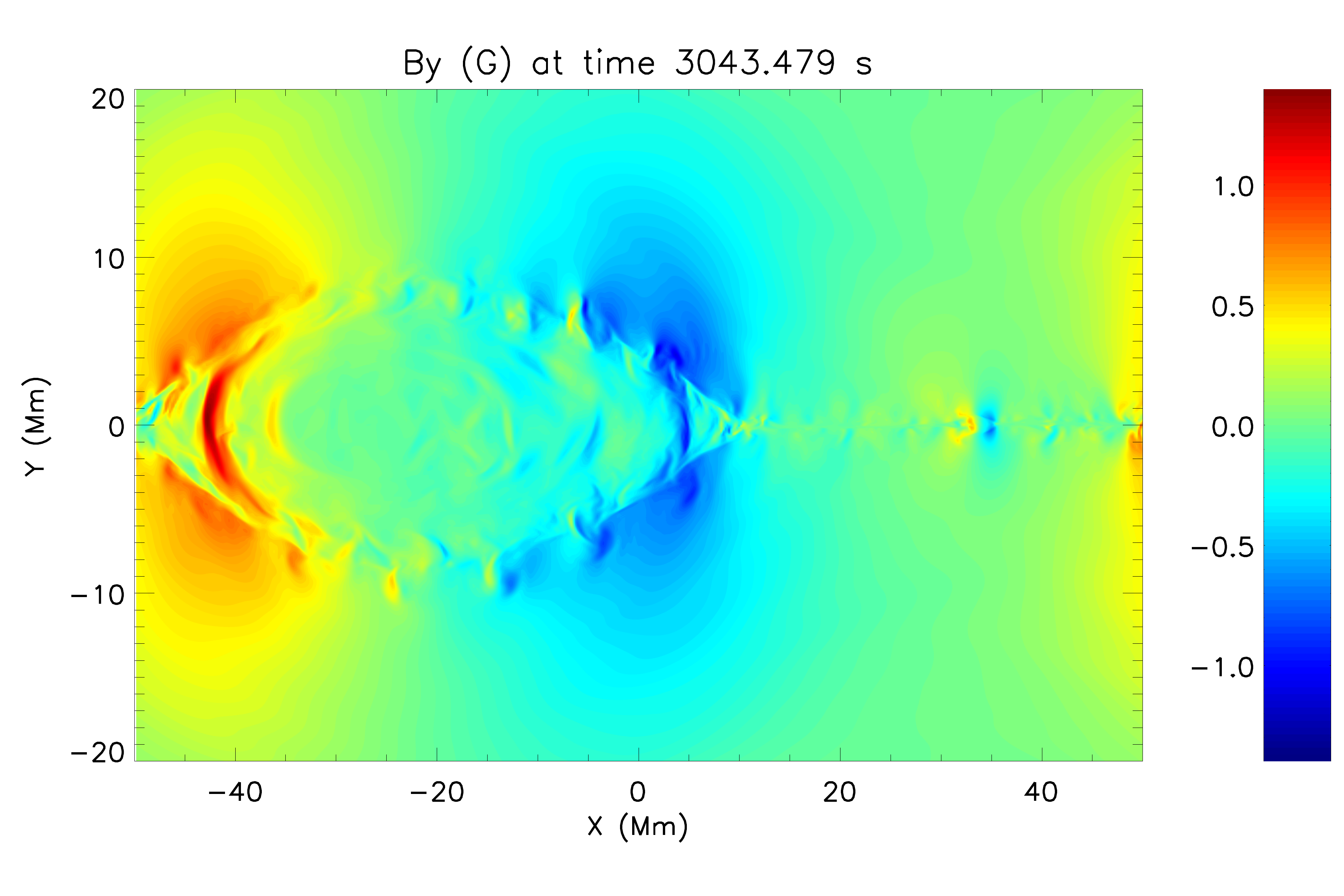} \\
\caption{$V_{x}$ and $B_{y}$ at $z=0$ for Simulation 3. \label{fig:lowshear_2D}}
\end{center}
\end{figure*}

Figure \ref{fig:energy} shows the internal energy gain (heating) as a function of time for all 6 simulations, normalized by the initial available free magnetic energy at the start of the simulation $\int_{V}{B_{x}^2dV}$. As previously mentioned, we solve the MHD equations non-conservatively, and do not feed the unrealistically large ohmic heating into the energy equation. Thus the only gain to the internal energy, and therefore the only plasma heating, is due to viscous heating at small gradients, modeled using the shock-viscosity formulation. {\color{black} As mentioned in \S\ref{sec:num} we have performed an estimation of the numerical loss of energy in the system}, and find that the magnitude of the energy lost/gained due to numerical effects is less than the kinetic energy for the linear stage. During the non-linear stage typically 20\% of the magnetic energy lost from the system is not recovered, due to numerical losses of the magnetic energy and the kinetic energy, i.e., numerical resistivity and numerical viscosity (the latter loss is also due to imperfect shock viscosity formulation).


In all 6 simulations, we follow the transition from linear into non-linear 3D tearing, and examine the  evolutionary path to the final state of each simulation. The path is dependent on the relative strengths of the available parallel and oblique modes, and their non-linear interaction.  The 6 simulations fall into two categories. The three long-sheet (large $L_x$) simulations (1,2,3), colored black, red, blue, respectively in Figure \ref{fig:energy}, show a qualitatively similar evolution of kinetic and internal energy, with a somewhat weak dependence on magnetic shear (the shear in the initial current sheet decreases from Sim. 1-3). The gradient of the internal energy gain curve is larger for stronger shear. In addition, the kinetic energy curves are similar for all three simulations, but are offset in time, with the time that the kinetic energy increases to significant values later for weaker shear. The three short-sheet simulations (4,5,6), colored magenta, green, and cyan, respectively in Figure \ref{fig:energy} have a much stronger dependence on shear (shear decreases from Sim. 4-6). Only Sim. 4 shows any significant heating, and any significant non-linear kinetic energy increase. Note that the kinetic energy scale in Figure \ref{fig:energy} is logarithmic. We therefore analyze Simulations 1-3 and 4-6 separately to ascertain the observed dependence of energy release on magnetic shear and current sheet length. We define $\lambda_{x,max}$ as the wavelength of the fastest growing mode that would be present in an infinite system, based on the theoretical growth rates above.

\subsection{Regime 1: $L_{x} > \lambda_{x,max}$}

As discussed in \S\ref{sec:validate}, the dominant tearing mode in Simulations (1-3) is a parallel mode ($n=0$) with $m>1$, such that this mode has multiple wavelengths of $\lambda_{x,max}$ inside the $x$ domain.  This results in multiple tearing islands forming along the current sheet. Figures \ref{fig:highshear_2D} and \ref{fig:lowshear_2D} show the non-linear evolution of $V_{x}$ and $B_{y}$ in Simulations 1 and 3. Note that the $y$ range is different for the $V_{x}$ and $B_{y}$ panels. Simulation 2 has been omitted for brevity, but shows qualitatively similar evolution. Although the dominant mode in Sims. 1-3 has multiple wavelengths per $x$ domain, as can be seen in the left panels (a),(c), and (e) of Figure \ref{fig:theory_rates}, there are modes with significant growth rates for lower $m$.
The non-linear interaction of the modes leads to coalescence of the islands (middle panels of Figures \ref{fig:highshear_2D} and \ref{fig:lowshear_2D}), and eventually one large island which saturates. One can think of the outflow jets from the long-wavelength modes forcing together the islands of the faster growing short-wavelength modes. The flows associated with coalescing islands generate velocity gradients leading to viscous heating, which results in the increase in internal energy seen in Figure \ref{fig:energy}. 
This heating saturates, due to the fact that the island consumes the available shear field and cannot grow any further, as shown in the bottom panels of Figure \ref{fig:highshear_2D} and \ref{fig:lowshear_2D} (the region of shear is restricted to $-25 < y < 25$ Mm, Eq. 7).

Although the oblique modes are significant in Simulations 1-3, and their relative importance increases with magnetic shear, the evolution of the islands is dominated by coalescence into one large island for all 3 Simulations, such that there is very little dependence on magnetic shear of the general evolution of the system, and ultimately only a weak dependence of the coronal heating due to the reconnecting islands on the magnetic shear. Figure \ref{fig:theory_rates} shows that the growth rates of some oblique modes are only modestly slower than the fastest parallel modes, but the difference in mode amplitudes becomes very large during the roughly 10 e-folding times that span the linear phase.

\begin{figure*}
\begin{center}
    \includegraphics[width=0.4\textwidth]{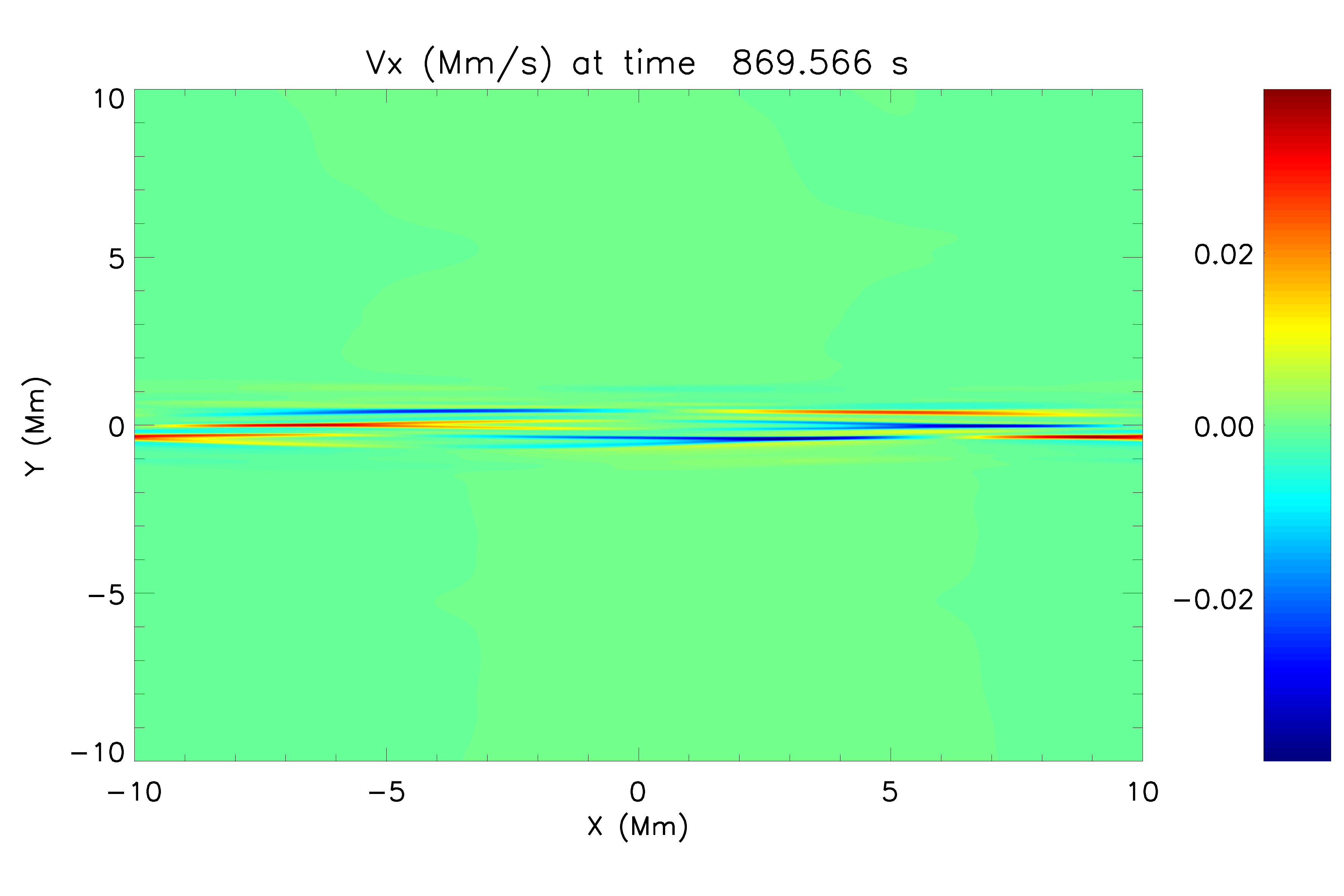}
    \includegraphics[width=0.4\textwidth]{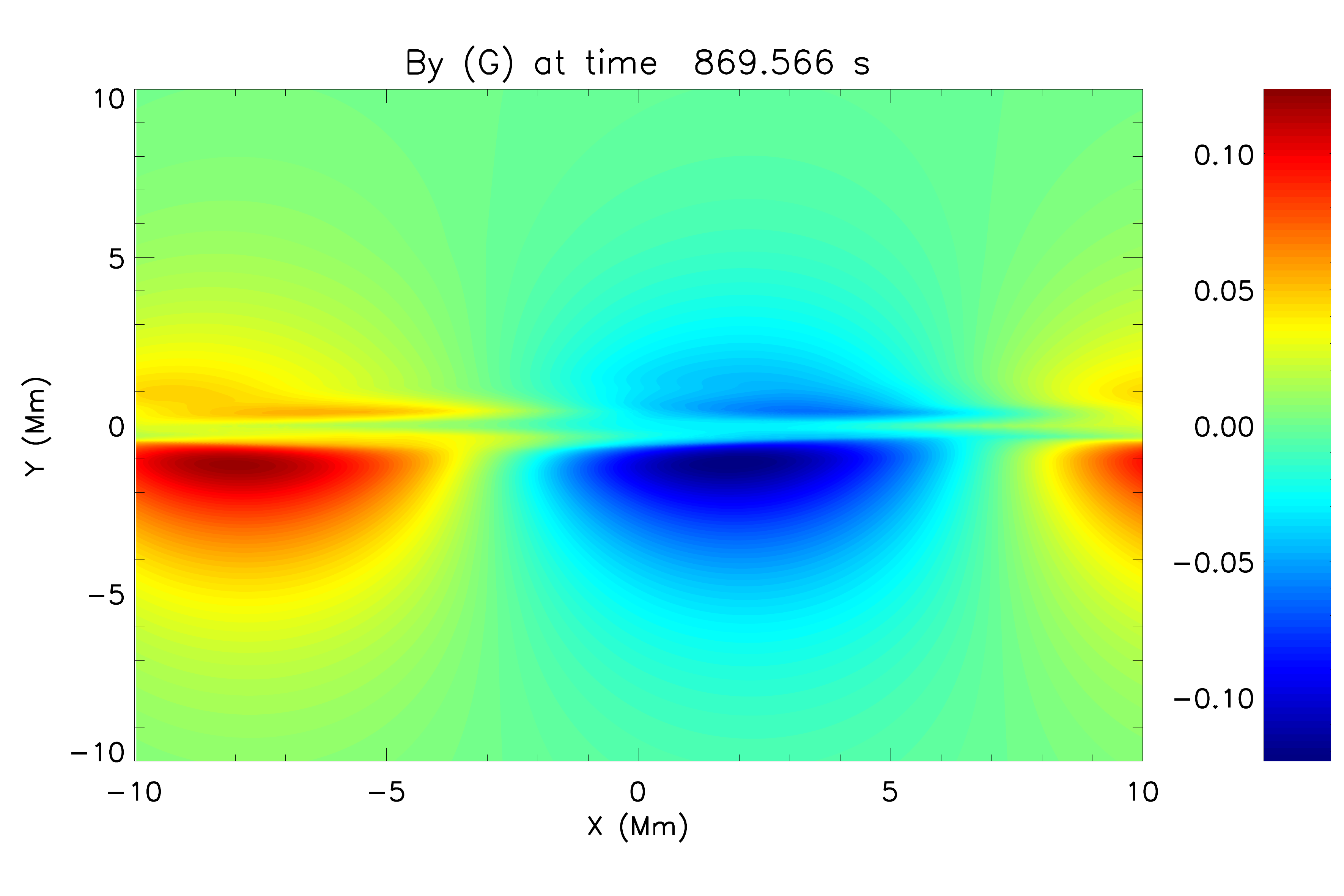}  \\
     \includegraphics[width=0.4\textwidth]{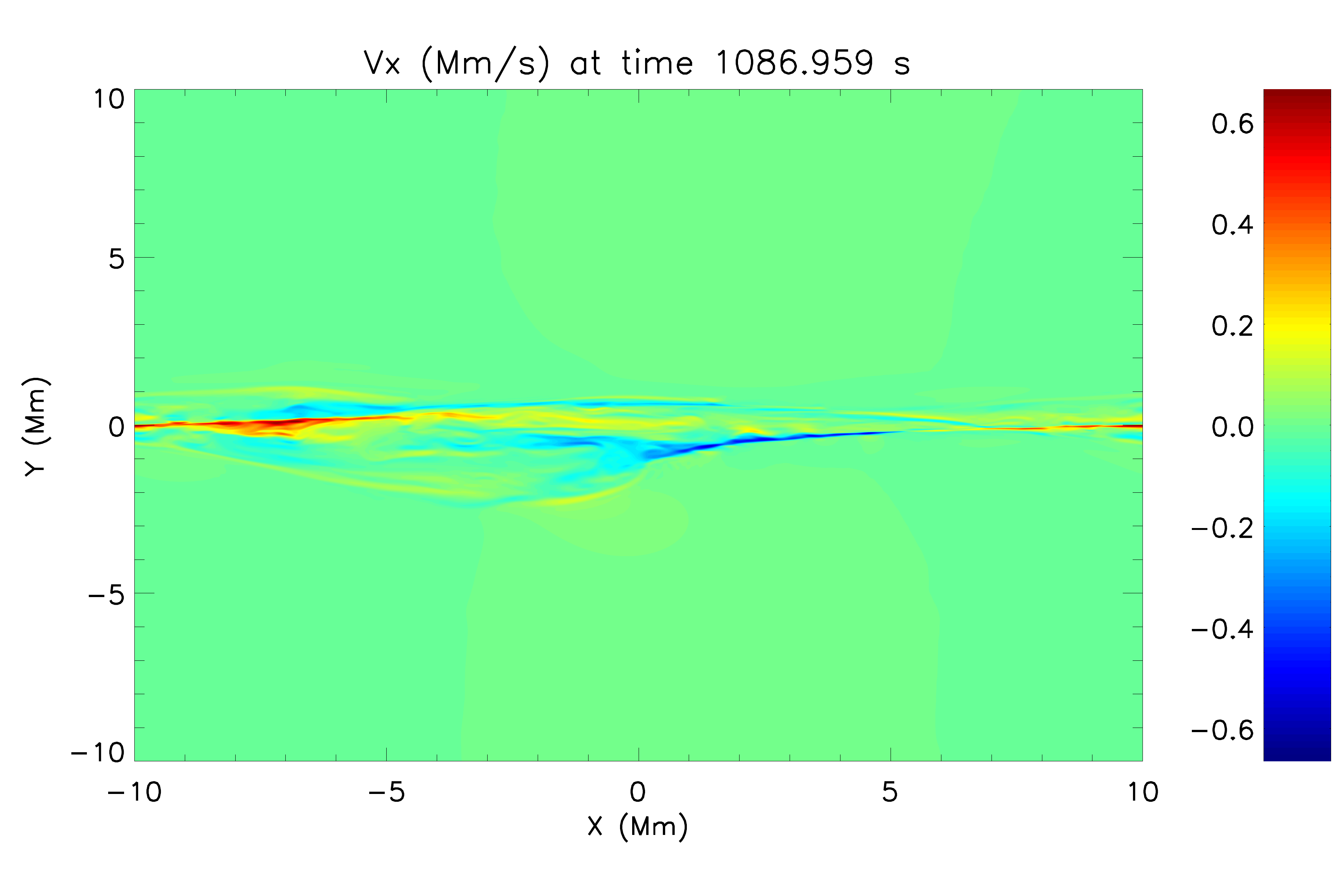}
     \includegraphics[width=0.4\textwidth]{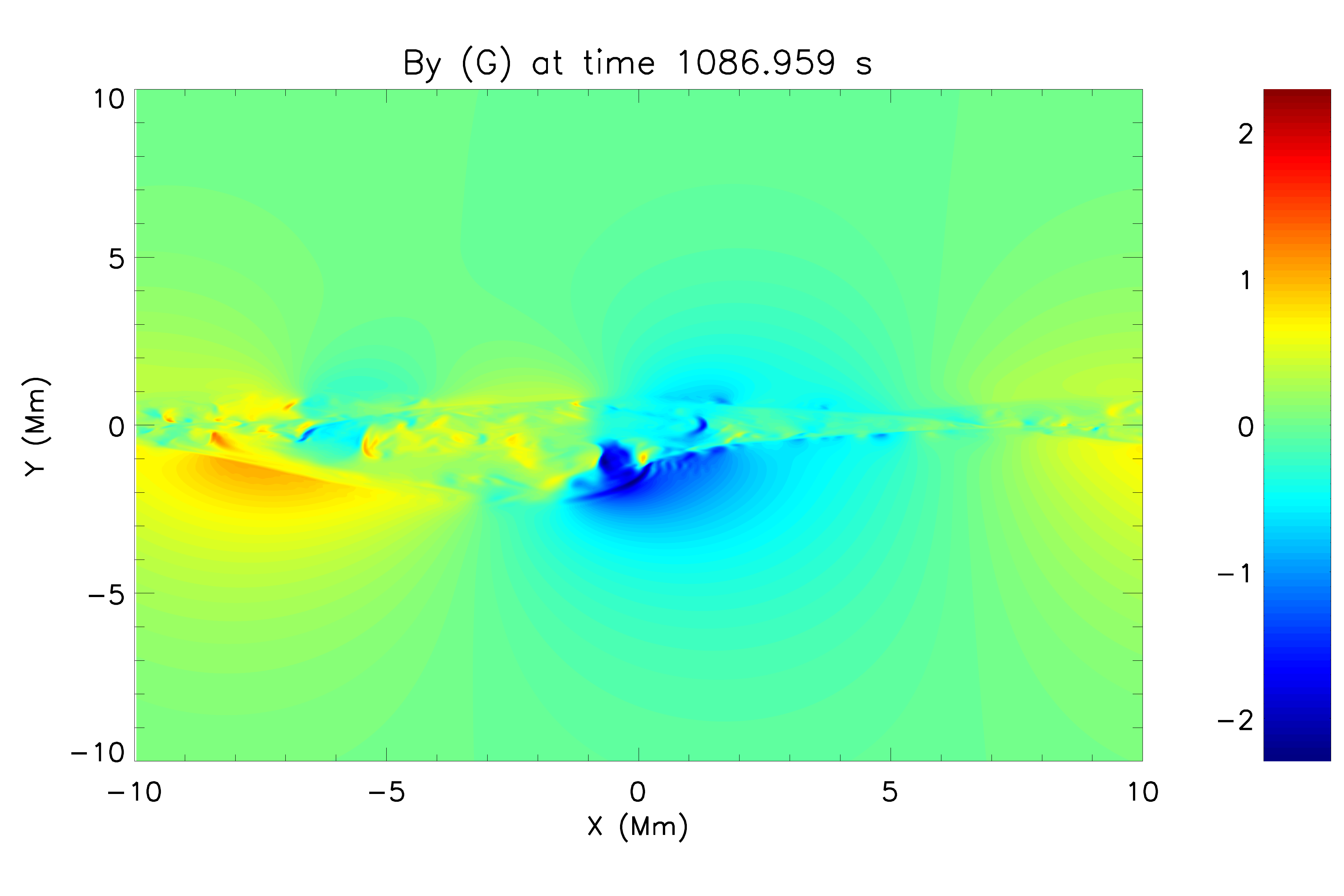} \\
    \includegraphics[width=0.4\textwidth]{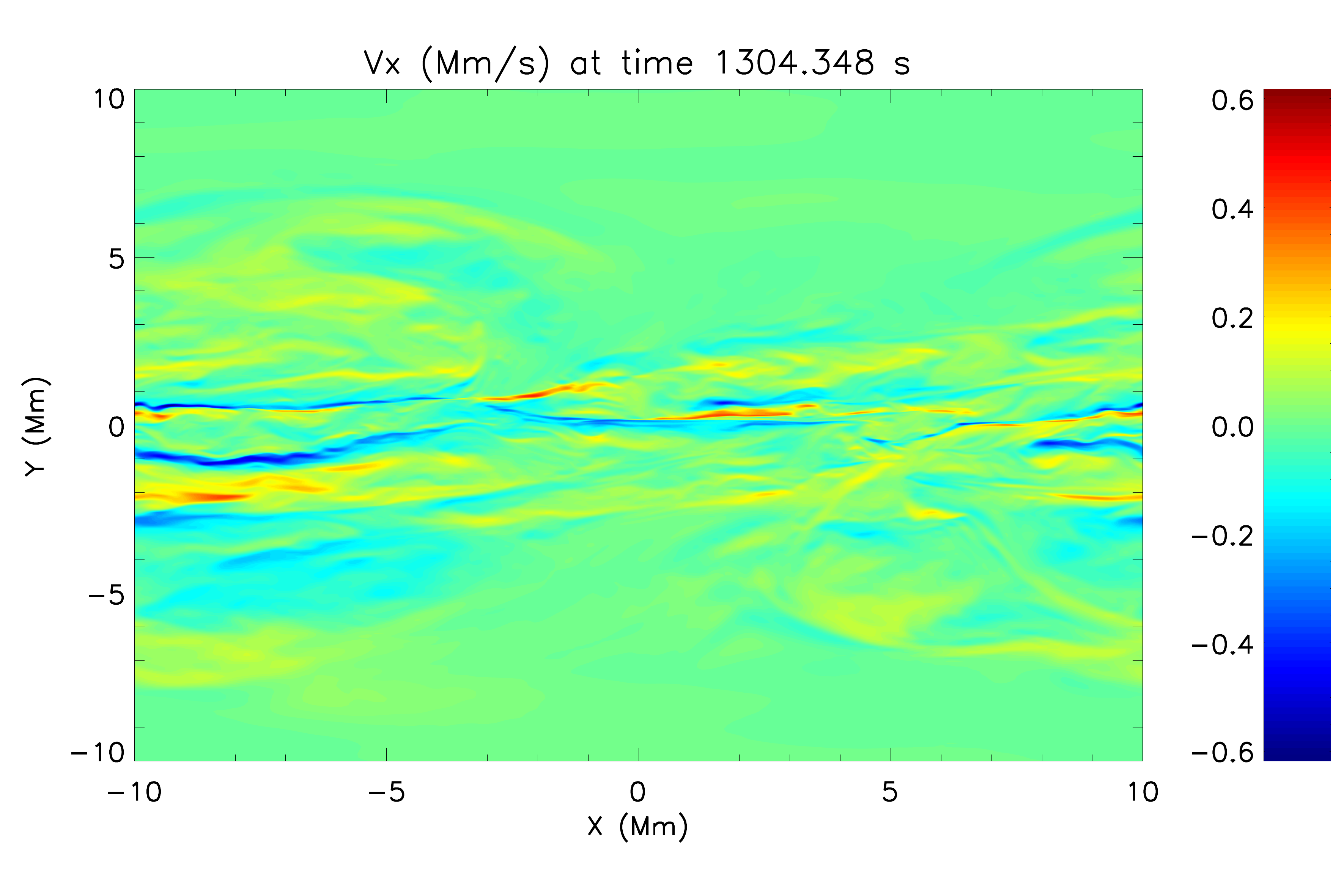}
    \includegraphics[width=0.4\textwidth]{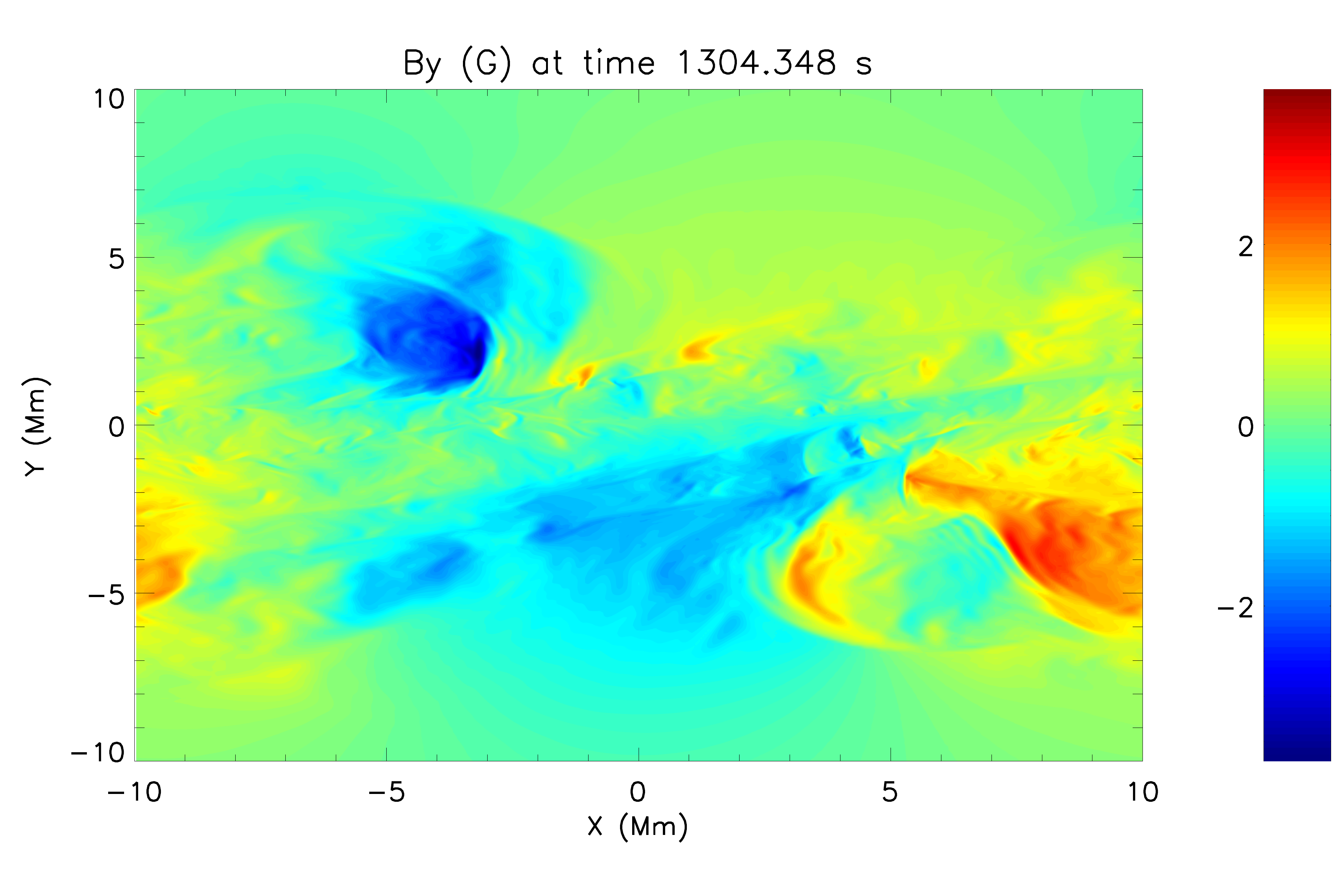}
    \caption{$V_{x}$ and $B_{y}$ at $z=0$ for Simulation 4.
    \label{fig:highshear_lowLx_2D}}
    \end{center}
    \end{figure*}
\begin{figure*}
\begin{center}
   \includegraphics[width=0.4\textwidth]{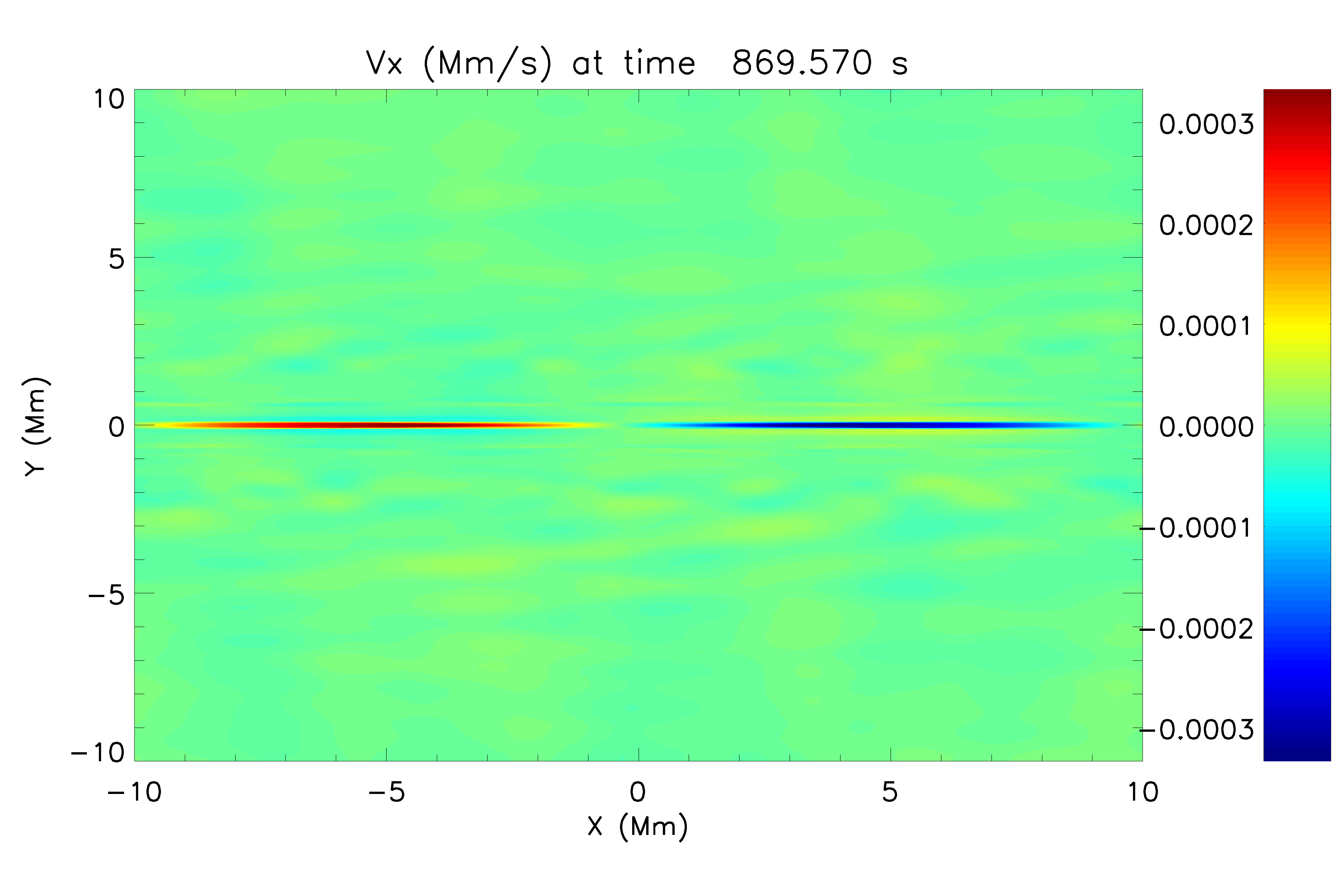}
    \includegraphics[width=0.4\textwidth]{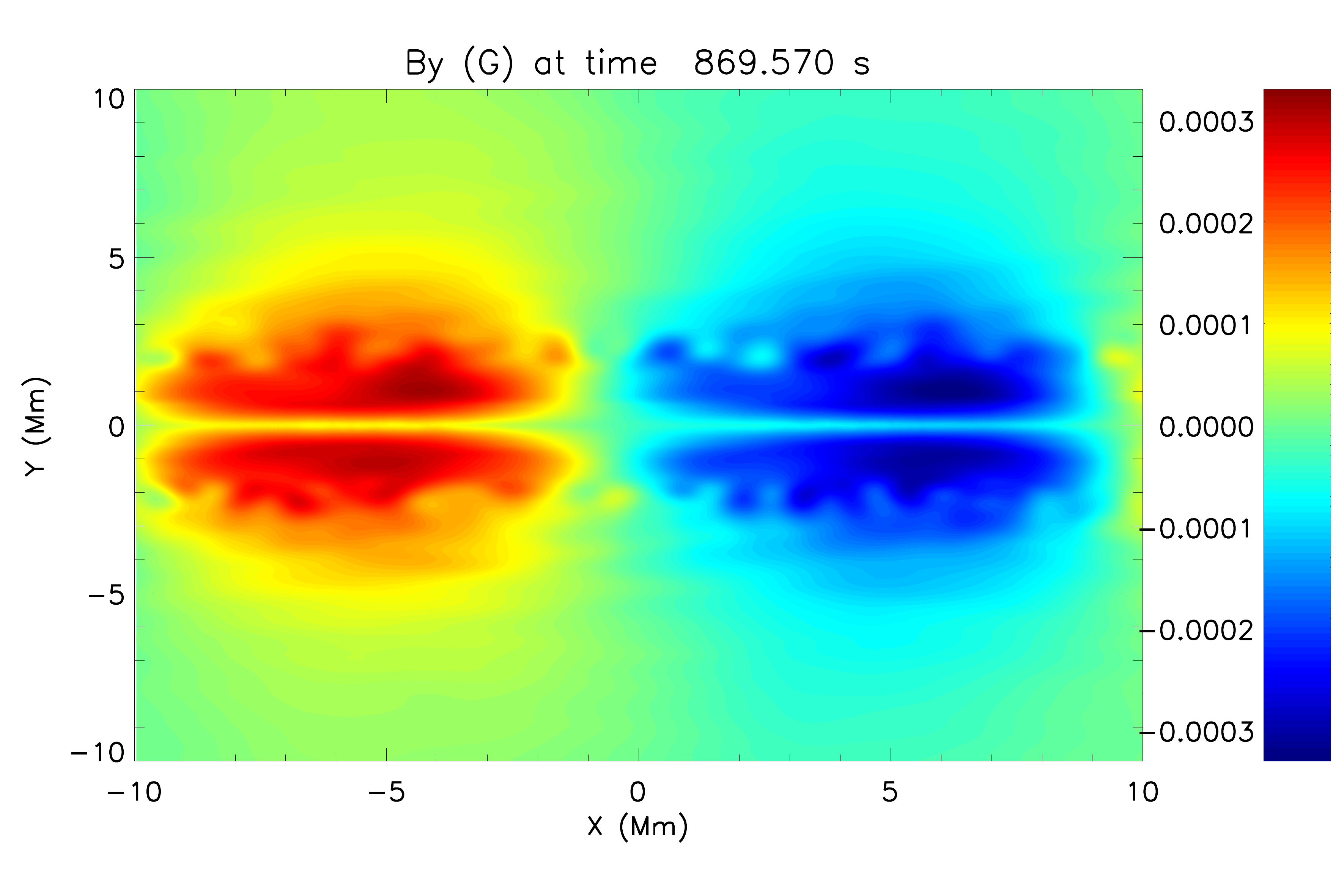}  \\
     \includegraphics[width=0.4\textwidth]{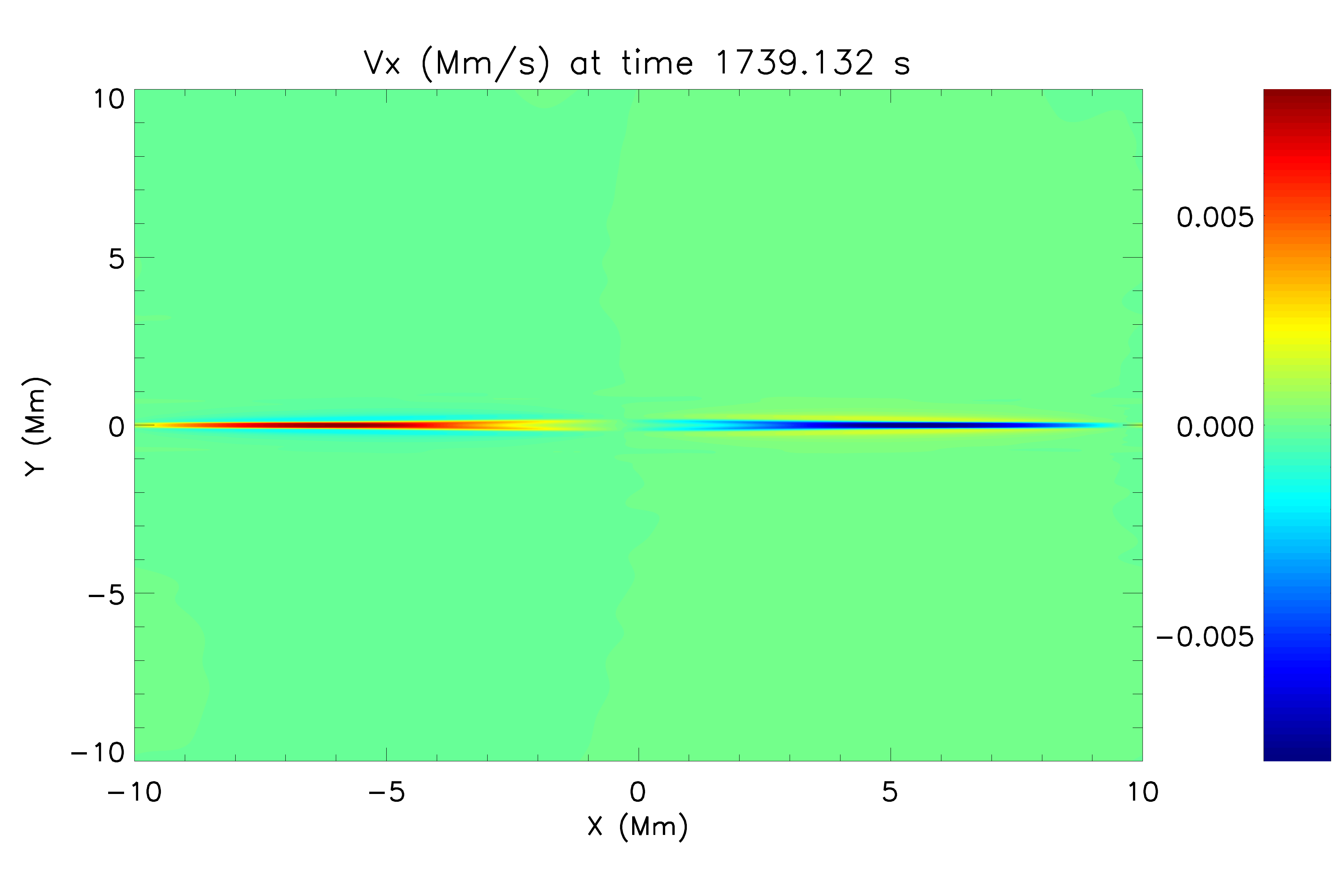}
     \includegraphics[width=0.4\textwidth]{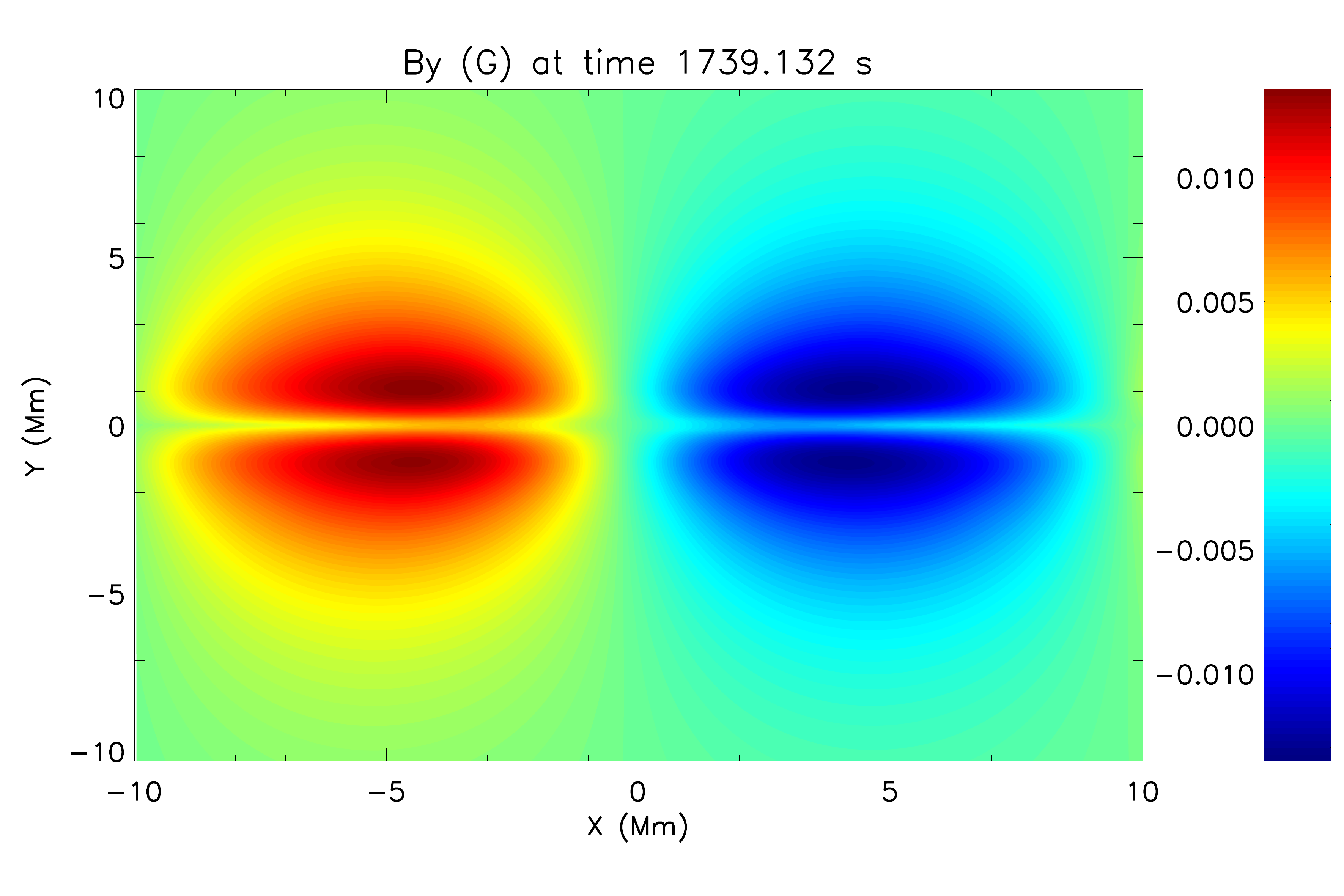} \\
    \includegraphics[width=0.4\textwidth]{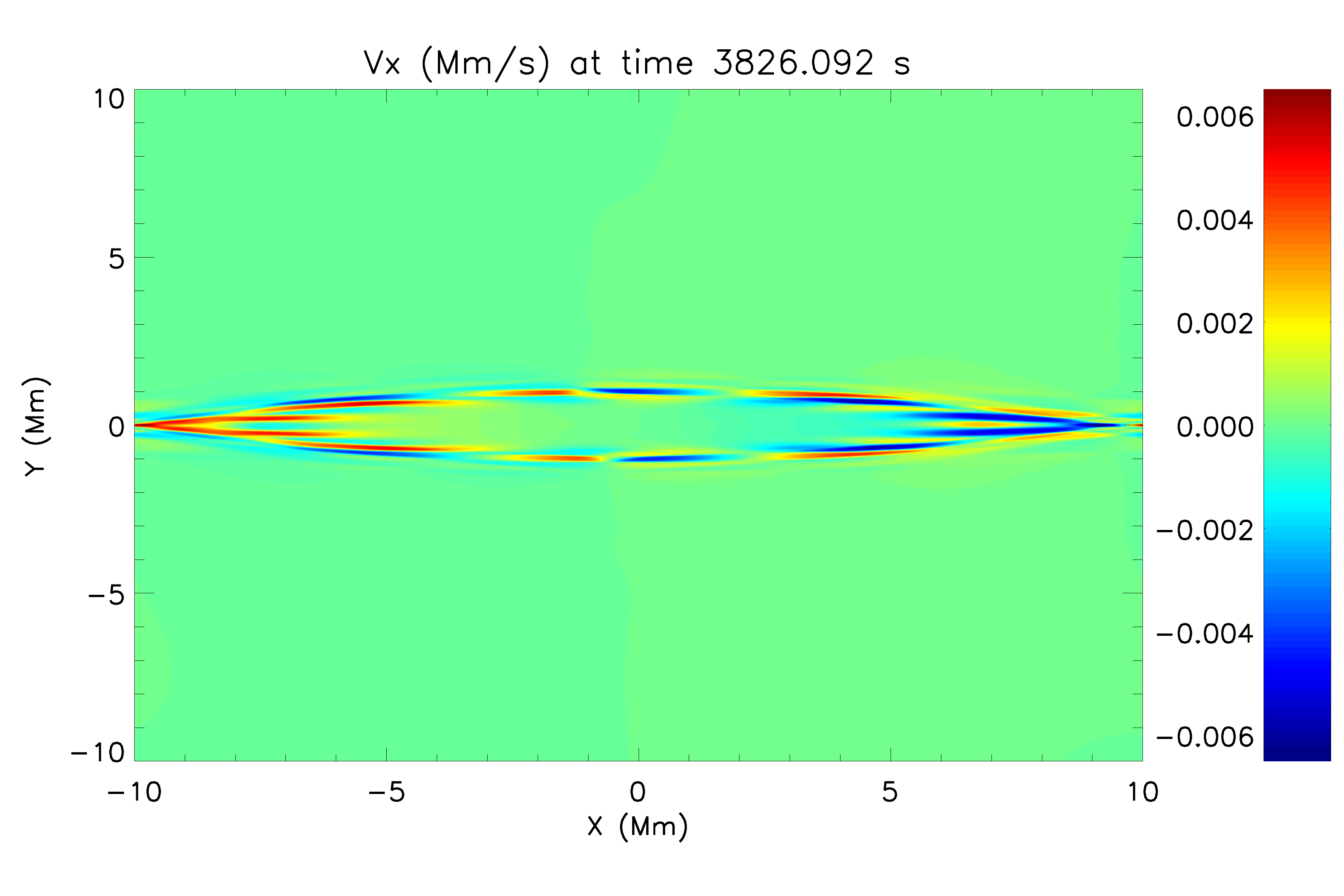}
    \includegraphics[width=0.4\textwidth]{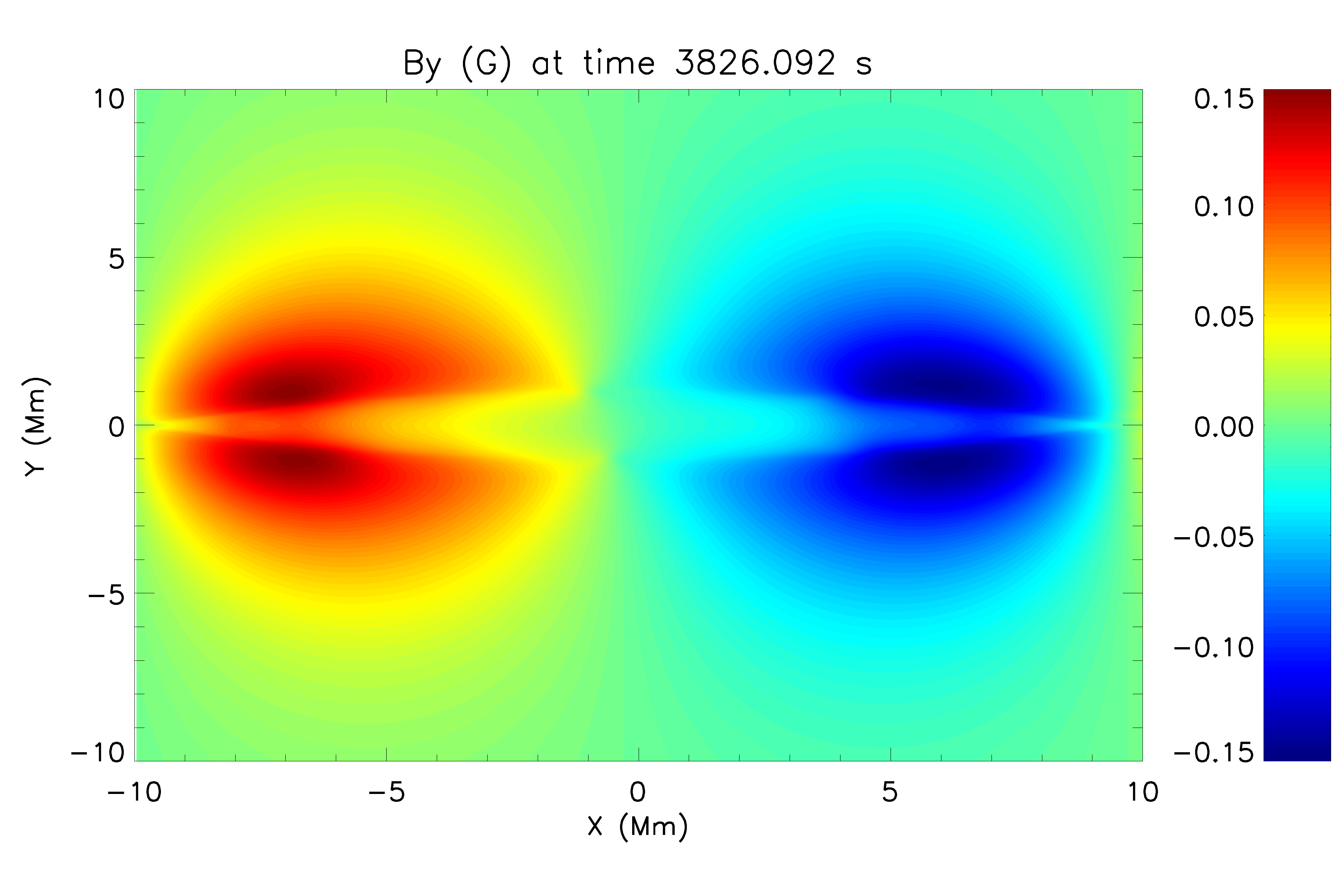}
   \caption{$V_{x}$ and $B_{y}$ at $z=0$ for Simulation 6.
    \label{fig:lowshear_lowLx_2D}}
  \end{center}
\end{figure*}

\subsection{Regime 2: $L_{x} < \lambda_{x,max} $}

In contrast to Simulations 1-3, Simulations 4-6 have dominant modes with $m=1$, also parallel, and therefore are in the regime where $L_{x} < \lambda_{x,max}$. Since there are no sub-harmonics of larger wavelengths, no coalescence can occur, as also emphasized by  \citet{Dahlburg_2002} and \citep{Onofri_2004}. This leads to a fundamentally different evolution of Simulations 4-6, and a very different dependence of coronal heating on magnetic shear. Figures \ref{fig:highshear_lowLx_2D} and \ref{fig:lowshear_lowLx_2D} show the evolution of $V_{x}$ and $B_{y}$ at $z=0$ for Simulations 4 and 6, with Simulation 5 omitted for brevity as it is in some sense intermediate between Sims. 4 and 6. We would note, however, that Sim. 5, like Sim.6, releases a far smaller fraction of the available free magnetic energy than does Sim. 4 (Fig. \ref{fig:energy} ).

 We consider first the strong shear case, Simulation 4. As indicated in Fig. \ref{fig:theory_rates} (b), the (1,1) oblique mode initially grows nearly as quickly as the (1,0) parallel mode. As a consequence, the two modes become nonlinear together, maintaining a similar amplitude. Once nonlinear, they interact strongly. This interaction is very complex because the flux tubes produced by the modes have different orientations. A non-laminar state develops and significant amounts of magnetic energy are released to heat the plasma. Note that in this case, the large scale structure of the current sheet is not disrupted, even though a significant amount of shear energy is converted to plasma heating. We note that a smaller fraction of the available free energy is released in Simulation 4 than in Simulations 1-3. This is because the processing of sheared field in Simulation 4 does not extend outside of the current sheet, as it does when a large island is formed from coalescence. If this holds true for actual solar current sheets, which are much thinner, then some mechanism of maintaining the reconnection, i.e., bringing external sheared field into the current sheet, would be needed to release truly large amounts of energy.

 The situation is much different in the weak shear case, Simulation 6. The (1,0) parallel mode grows substantially faster than the fastest growing oblique mode, (3,1), as indicated in Fig. \ref{fig:theory_rates}, panel (f). By the time the parallel mode saturates, it has modified the initial equilbrium enough that the oblique mode is quenched at small amplitude. (The oblique mode is an unstable mode of the initial equilibrium, not the modified equilibrium, and so is different from the secondary instability discussed by \citet{Dahlburg_2005}.) Minimal magnetic energy is released. In the bottom left panel of Fig. \ref{fig:lowshear_lowLx_2D} it appears that new unstable modes have started to develop around the edges of the saturated island in the modified equilibrium. We do not anticipate that they will grow to substantial size. Note that the color scales are much different for Sim. 6 than for Sim. 4. Velocities peak at about 500 km/s at the final time in Fig. \ref{fig:highshear_lowLx_2D}, but only about 7 km/s in Fig. \ref{fig:lowshear_lowLx_2D}. They also peak at about 500 km/s at the final time in the long current sheets of Figs. \ref{fig:highshear_2D} and \ref{fig:lowshear_2D}. 

The middle and lower panels of Figures  \ref{fig:highshear_lowLx_2D} and \ref{fig:lowshear_lowLx_2D} highlight how much Simulations 4 and 6 diverge later in the non-linear phase. For Simulation 6, with weaker oblique modes, the parallel mode dominates, creating a single tearing island which saturates.  
For Simulation 4, the interaction of the equally strong oblique and parallel modes leads to a rapid disruption of the current sheet, non-laminar flows, and this results in a large increase in kinetic energy and ultimately plasma heating, as seen in the energy curves for Simulation 4 (magenta lines) in Figure \ref{fig:energy}. This divergence of behavior, which is a result of stronger oblique modes for high shear, exacerbated over many growth times into the non-linear regime, results in a strong dependence of heating on magnetic shear, and suggests that magnetic shear is a good candidate for the switch-on parameter for rapid coronal reconnection and heating in this particular parameter space.

\section{Discussion}
\label{sec:discussion}
We have performed a small but relevant parameter study of the onset of magnetic reconnection and heating in the solar corona via the 3D tearing instability, focusing on the parameters of magnetic shear across the 3D current sheet and the current sheet length. We chose two values of $L_{x}$ and three values of magnetic shear. The shear values were chosen to encompass the theoretical Parker angle, above which it was suggested, based on coronal energy balance considerations, that sheared fieldlines would reconnect and release magnetic energy (see discussion in \citet{Klimchuk_2015}).

For our chosen parameters, we find two regimes. The first is the large $L_{x}$ regime, where the dominant (fastest growing) linear mode has sub-harmonics (smaller $m$, therefore larger wavelength modes). The second is the smaller $L_{x}$ regime, where there are no sub-harmonics of the dominant mode. In all cases, the dominant mode is a parallel mode centered on the sheet. Oblique modes can nonetheless be critically important. We find three evolutionary paths.

\begin{itemize}

\item  If subharmonics exist (e.g., Sims. 1-3), the tearing islands  coalesce to ultimately form one large island.
	It happens more quickly for large shear, and the heating rate is weakly proportional to the magnetic shear.
	Coalescence  is rapid and dominates any other effects associated with oblique modes. In our simulation, the single island grows until its cross-sheet dimension is comparable to the width of the shear zone, and
 10-30\% of the available free magnetic energy in the system is converted into plasma heating. In more realistic environments, the island may continue to grow until it encompasses the length of the sheet and takes on a roughly circular shape. This is the minimum magnetic energy state.
\item If no subharmonics exist and the shear is small (e.g., Sims. 5 and 6), the fastest (1,0) parallel mode significantly outpaces the slower oblique modes and saturates at modest but nonlinear amplitude while the oblique modes are still small. The oblique modes stop growing in the modified equilibrium, and minimal energy is released.
\item If no subharmonics exist and the shear is large (e.g., Sim. 4), at least one oblique mode grows nearly as quickly as the dominant (1,0) parallel mode. The modes reach nonlinear  amplitude together and  interact strongly,  leading to a turbulent disruption of the current sheet and large release of magnetic energy. 
\end{itemize}

Our results and their interpretation are consistent with the long current sheet ($ L_{x} > \lambda_{x,max} $) simulations of \citet{Onofri_2004} and \citet{Landi_2008}, and the short current sheet ($L_{x} < \lambda_{x,max}  $) simulations of \citet{Dahlburg_2005} and \citet{Dahlburg_2006}. We attribute the rapid and large energy release in short sheets with large shear to the nonlinear interaction of different parallel and oblique modes that grow together from small amplitude, as discussed previously by \citet{Huang_2016}. This interpretation differs from that of Dahlburg and colleagues, starting with \citet{Dahlburg_2002}, who propose that a secondary instability sets in after the initial tearing, consisting only of parallel modes, saturates at a modest level. This secondary instability is attributed to kinking. Further analysis is required to determine which interpretation is correct.

Our result that the evolution of long sheets is dominated by coalescence for all levels of shear, and not by secondary instability, was also found by \citet{Onofri_2004} and \citet{Landi_2008}. This appears to be inconsistent with \citet{Dahlburg_2002}, who found that secondary instability proceeds more quickly than coalescence. We suggest that the outcome of the latter simulations may be determined by the specific perturbations that were applied to initiate the evolution in their studies. Two perturbations were applied together:  1. an oblique mode having a wavelength in the shear direction equal to that of the fastest growing parallel mode; and 2. a parallel mode that is the first subharmonic of the fastest growing parallel mode. The fastest growing parallel mode was not itself introduced at the outset. It appears that a broad spectrum of random perturbations that excites all modes, as in our investigation, leads to the dominance of coalescence over secondary instability. This is of course more realistic. In addition, our choice of relatively small resistivity and hence large Lundquist number, and the relatively strong guide field, leads to a system where the oblique modes are of comparable strength to the parallel modes, leading to both coalescence and oblique interaction occurring depending on the choice of magnetic shear in the current sheet.

All simulations with coalescence (Sims. 1-3) eventually produce one large island, or flux tube, whose radius is determined by the smallest relevant dimension in the plane of reconnection (shear and cross-sheet directions). In our case, the appropriate cross-sheet dimension is the region of sheared field, which is half the box size. Depending on the degree of shear, this mega-flux tube can be kink unstable. Simple geometric considerations can be used to estimate the amount of twist in the guide field direction and thereby predict stability or instability. All three of our long-sheet simulations are predicted to end with a stable flux tube, as observed. The non-force-free equilibria with and without a guide field in \citet{Landi_2008} are also correctly predicted to produce stable and unstable flux tubes, respectively. We note that stability to kinking depends on the number of turns of twist along the axis of the flux tube. Therefore, the dimension of the box  in the guide field direction (or of the current sheet in a real physical system) is also an important parameter.



{\color{black}
It is worth contextualizing our parameter regime with respect to the two major consequences of magnetic reconnection in the solar corona mentioned in the introduction; the heating of the corona itself and eruptive flare/CMEs in the corona. Before we discuss each scenario separately,  we show here a rough estimate of the wavelength of the most unstable parallel mode given real solar parameters of $V_{a} = 10^{6}$ m/s, $D=\eta/\mu_{0} = 1 \textrm{m}^{2}/\textrm{s}$ and a chosen tearing mode growth rate of $\tau_{max} = 10$ s. Using the estimates of the most unstable wavenumber and the maximum growth rate of the 3D tearing mode from \citet{Baalrud_2012}: $k_{max}a = S_{a}^{-1/4}$, $\tau_{a}/\tau_{max} = S_{a}^{-1/2}$ where $S_{a} = aV_{a}/D$ and $\tau_{a} = a/V_{a}$, the most unstable wavelength can be shown to be $\lambda_{max} = 2\pi V_{a}^{2/3} D^{1/6} \tau_{max}^{5/6}$ which for the parameters above is approximately 430 km.
 
 The magnetic field of the corona is comprised of a multitude of topologically distinct elemental flux tubes that are rooted in the high-beta plasma at the solar surface. Random motions associated with photospheric convection tangle and twist the tubes, resulting in current sheets at the boundaries where they press against each other in the corona. The length of the sheets in the shear direction is comparable to the flux tube diameter, which is of order 100 km \citep{Klimchuk_2015}. These current sheets, where coronal heating takes place, are therefore in the short sheet regime, $L_x < \lambda_{max}$. Since a sizable shear angle is required for significant energy release in this regime, this offers a natural explanation for the ?Parker angle inferred by comparing observed coronal energy losses with the Poynting flux associated with photospheric driving. This suggests that 
magnetic shear is a candidate for a switch-on parameter, as we see a strong dependence of the reconnection and heating on the magnetic shear in this regime. Further studies, using the parameters chosen here, and similar MHD simulations but with driven boundary conditions representing the slow shearing and twisting of fieldlines, will test this hypothesis.}

For situations such as eruptive flares and CMEs, one should expect very long current sheets, and therefore that the system should lie in the regime $L_{x}>\lambda_{x,max}$, where there are multiple parallel modes and hence many interacting islands. In our study, we see a weak dependence of the reconnection-driven plasma heating on magnetic shear, so in these scenarios, we suggest that another parameter would provide the switch-in nature. As discussed in \S\ref{sec:intro}, it has been suggested that thinning current sheets become unstable to tearing well before reaching Sweet-Parker widths. Well resolved numerical MHD studies of thinning 3D current sheets in coronal conditions are required to investigate this further.


\section{Acknowledgements}

We thank the many members of the solar physics and magnetic reconnection community with whom we interacted for useful and illuminating discussions. {\color{black} In particular we thank the following for discussions which improved the study in this manuscript: Paul Cassak, Yi-Min Huang, Scott Baalrud, Marco Velli, Anna Tenerani, Kalman Knizhnik,  Russell Dahlburg, Spiro Antiochos, Richard DeVore, Nuno Loureiro, and Rebekah Evans}. 
Numerical simulations were performed on NASA supercomputing, using NASA Heliophysics supported funds. This work was funded by NASA's Heliophysics Supporting Research Program, and by NASA's Heliophysics Internal Scientist Funding Model (competitive work package) program at Goddard Space Flight Center.

\bibliographystyle{apj}

\end{document}